\documentclass[lettersize,journal]{IEEEtran}
\usepackage{amsmath,amsfonts,bm}
\usepackage{algorithmic}
\usepackage{algorithm}
\usepackage{array}
\usepackage{textcomp}
\usepackage{stfloats}
\usepackage{url}
\usepackage{verbatim}
\usepackage{graphicx}
\usepackage{cite}
\usepackage{booktabs}
\usepackage{multirow}
\usepackage{tipa}
\usepackage[OT2,OT1,T1]{fontenc}
\usepackage{hyperref}  
\usepackage[usenames,dvipsnames]{color}
\hypersetup{colorlinks=true,pdfborder=0 0 0,citecolor=blue,linkcolor=blue,urlcolor=blue}
\usepackage{amssymb}
\usepackage{color}

\begin{document}
\newcommand{\RI}{\textcolor{black}}

\title{Whistle: Data-Efficient Multilingual and Crosslingual Speech Recognition via Weakly Phonetic Supervision}

\author{Saierdaer Yusuyin, Te Ma, Hao Huang, ~\IEEEmembership{Member,~IEEE}, Wenbo Zhao, Zhijian Ou, ~\IEEEmembership{Senior Member,~IEEE}

\thanks{This work was supported by National Science and Technology Major Project (2023ZD0121401), Guangxi Science and Technology Project (2022AC16002), National Natural Science Foundation of China (62466055), and a funding from TasiTech.
Corresponding author and principal investigator of this work: Zhijian Ou.}

\thanks{Saierdaer Yusuyin, Te Ma, Hao Huang are with the School of Computer Science and Technology, Xinjiang University, Urumqi 830046, China (e-mail: sar\_dar@foxmail.com; mate153125@gmail.com; huanghao@xju.edu.cn)}
\thanks{Wenbo Zhao is with the China Unicom (Guangdong) Industrial Internet Co., Ltd, Guangzhou 510555, China 
(e-mail: zhaowb19@chinaunicom.cn)}
\thanks{Zhijian Ou is with the Speech Processing and Machine Intelligence (SPMI) Lab, Department of Electronic Engineering, Tsinghua University, Beijing 100084, China, 
(e-mail: ozj@tsinghua.edu.cn)}
}



\maketitle

\begin{abstract}
There exist three approaches for multilingual and crosslingual automatic speech recognition (MCL-ASR) - supervised pretraining with phonetic or graphemic transcription, and self-supervised pretraining.
We find that pretraining with phonetic supervision has been underappreciated so far for MCL-ASR, while conceptually it is more advantageous for information sharing between different languages.
This paper explores the approach of pretraining with weakly phonetic supervision towards data-efficient MCL-ASR, which is called Whistle.
We relax the requirement of gold-standard human-validated phonetic transcripts, and obtain International Phonetic Alphabet (IPA) based transcription by leveraging the LanguageNet grapheme-to-phoneme (G2P) models.
We construct a common experimental setup based on the CommonVoice dataset, called CV-Lang10, with 10 seen languages and 2 unseen languages. 
A set of experiments are conducted on CV-Lang10 to compare, as fair as possible, the three approaches under the common setup for MCL-ASR. 
Experiments demonstrate the advantages of phoneme-based models (Whistle) for MCL-ASR, in terms of speech recognition for seen languages, crosslingual performance for unseen languages with different amounts of few-shot data, overcoming catastrophic forgetting, and training efficiency.
It is found that when training data is more limited, phoneme supervision can achieve better results compared to subword supervision and self-supervision, thereby providing higher data-efficiency.
To support reproducibility and promote future research along this direction, we release the code, models and data for the entire pipeline of Whistle at \url{https://github.com/thu-spmi/CAT/tree/master/egs/cv-lang10}.
\end{abstract}


\section{Introduction}
\IEEEPARstart{I}{n} recent years, deep neural network (DNN) based automatic speech recognition (ASR) systems have achieved significant progress, which are, however, data-hungry. A substantial amount of transcribed speech data are required for model training. There are more than 7,000 languages spoken around the world \cite{Ethnologue}, but due to the lack of training data, only a small fraction of them benefit from current ASR technology. An important challenge for the speech community is that we can develop ASR systems to new unsupported languages rapidly and at reasonable costs. Multilingual and crosslingual ASR (MCL-ASR) have been studied as an effective way to address this problem. 

In \emph{multilingual speech recognition}, training data for a number of languages, often referred to as seen languages, are merged to train a multilingual model, which can be used to recognize speech from all seen languages. The multilingual model can also serve as a pretrained model, which can be further finetuned for crosslingual speech recognition.
\emph{Crosslingual speech recognition} refers to recognizing utterances in a new language, which is unseen in training the multilingual model.
From machine learning perspective, such multilingual and crosslingual training can be regarded as performing multi-task learning and transfer learning, which promotes sharing of statistical strength.
The advantage is that the ASR performance for low-resource languages, both seen and unseen, can be improved, and the cost of system building and maintenance for multiple languages can be reduced as well.


The general concept of multilingual and crosslingual speech recognition has been applied for a long time, dating back to the time when GMM-HMM based classic models and then DNN-HMM based hybrid models are prevalent in ASR research, to name a few, e.g., in \cite{lu2013cross} and \cite{huang2013cross} respectively.
Recently, end-to-end models have emerged \cite{ctc, graves2012sequence,chorowski2014end}, which can be directly trained from phonetic or graphemic transcription, eliminating the first pass of producing HMM state alignment \RI{as used in DNN-HMM based hybrid models}.
For end-to-end models, the approach of pretraining followed by finetuning has attracted increasing interests and achieved good performance.
There are mainly two classes of pretraining methods, based on either self-supervised learning or supervised learning.
Self-supervised pretraining is conducted over unlabeled speech data from multiple languages for speech representation learning in general \cite{XLSR,XLS-R,MMS}. 
Supervised pretraining, by applying end-to-end models on multilingual labeled speech data, can be further divided into two sub-categories of research, which are contrasted by using different types of modeling units.
The first is grapheme-based or subword-based \cite{scalingMLASR,meta50,meta70,whisper}, which, collectively referred to as based on graphemic transcription (orthography), creates a shared token set across multiple languages, e.g., using 10K sentence pieces \cite{meta50}. 
The second trains end-to-end models on phonetic transcriptions \cite{li2020universal,joinAP,tachbelie2022multilingual,xu22b_interspeech,saier}, which usually utilizes International Phonetic Alphabet (IPA) symbols to create a (nearly-)universal phone inventory, e.g., using 187 phones \cite{li2020universal}.


Intuitively, the key to successful multilingual and crosslingual recognition is to optimize information sharing during multilingual training and maximize the knowledge transferring from a well trained multilingual model to the model trained for recognizing utterances in a new language \cite{joinAP}. 
Taking this perspective, we can examine the pros and cons of the three approaches - \emph{supervised pretraining with graphemic transcription or phonetic transcription, and self-supervised pretraining}, which is detailed in Section \ref{sec:related_work}. \RI{And there have two interesting research questions (RQs).}

\RI{The first question is about the comparison between phonetic supervision and graphemic supervision in pretraining for MCL-ASR.}
While requiring pronunciation lexicons, pretraining with phonetic supervision is more advantageous for information sharing between different languages. For phonetic supervision, IPA symbols include enough symbols to represent the fundamental sounds of all languages, and sounds in different languages share these phonetic representations \cite{fromkin2007introduction}.
In contrast, graphemes and subwords are in fact from writing systems of languages (orthography), not for describing and distinguishing all the sounds in human language throughout the world, which is exactly phonetic transcription does.
Creating a graphemic token set from multiple languages for supervision is non-trivial and delicately affects ASR performance; until recently, tokenization strategy is still under investigation and needs a balance between granularity and ASR performance \cite{meta70}; adding new languages for crosslingual recognition further complicates the design of tokenization. 
Besides the above theoretical analysis of supervised pretraining with graphemic transcription and phonetic transcription, an interesting research question is about empirical comparison.
It has been empirically found that compared to learning with graphemic supervision, learning with phonetic supervision performs equally strong and tends to be more {data-efficient} in monolingual ASR \cite{zeineldeen2020systematic,CTCCRF_IC19,an2020cat,zheng2021advancing}. 
But to the best of our knowledge, there have been no solid experiments to study which approach is better or if they yields similar results for MCL-ASR, when evaluated in a common experimental setup (Research Question 1, referred to as RQ-1).


\RI{The second interesting research question} is to compare supervised pretraining and self-supervised/unsupervised pretraining. Basically, we agree with the comments in \cite{whisper}. Current pretrained models for speech such as based on wav2vec 2.0 \cite{baevski2020wav2vec} aim to learn speech representation in general over unlabeled data; They mostly are encoder-only and thus lack an equivalently performant decoder, which requires at least adding a classifier layer and supervised finetuning over labeled data even for seen languages. These comments, presumably, are suited to comparing self-supervision to both graphemic supervision \cite{whisper} and phonetic supervision (our work).
These being said, to the best of our knowledge, there have been no strict experiments to study which approach is better or if they yields similar results for MCL-ASR, when evaluated in equal settings (Research Question 2, referred to as RQ-2).

\RI{Remarkably, in evaluating and comparing different pretraining approaches to answer the above research questions for MCL-ASR, data-efficiency is an important aspect.
Data-efficiency can refer to many things\footnote{\url{https://en.wikipedia.org/wiki/Data_efficiency}}, and in this paper, it mainly refers to achieving better performance with the same amount of data.
In particular, data-efficiency in MCL-ASR entails: 1) efficient use of pretraining data (referred to as pretraining data-efficiency or multilingual data-efficiency); and 2) efficient use of limited target language finetuning data such as 1 or 10 hours (referred to as finetuning data-efficiency or crosslingual data-efficiency).
The former evaluates the performance of a pretrained, multilingual model in recognizing seen languages, while the latter measures the performance of finetuned models in recognizing unseen languages.}


\vspace{+1mm}
\RI{In this paper, we present our effort to answer the above two research questions. Our main contributions are as follows.}
\RI{First,} we construct a common experimental setup based on the CommonVoice dataset, called CV-Lang10, to evaluate multilingual and crosslingual speech recognition, with 10 seen languages and 2 unseen languages, measuring both phoneme error rate (PER) and word error rate (WER). A set of experiments are conducted on CV-Lang10 to compare, as fair as possible, the three approaches under the common setup - supervised pretraining with graphemic transcription or phonetic transcription, and self-supervised pretraining for MCL-ASR.
\RI{It is found in our experiments that phonetic supervision obtains better multilingual data-efficiency than graphemic supervision\footnote{\RI{Note that a multilingual model from self-supervised pretraining is essentially just an encoder and thus alone cannot be directly applied in recognizing even seen languages. So for the self-supervision approach, we only examine crosslingual data-efficiency and do not consider multilingual data-efficiency. See more discussions at the end of Section \ref{sec:related_work_C}.}};
and compared to both graphemic supervision and self-supervision, phonetic supervision excels in crosslingual data-efficiency.}

\vspace{+1mm}
\RI{Second,} to address the problem of requiring phonetic transcription for phonetic supervision, we note that phonetic resources and tools have been steadily developed over these years and are easily accessible, including grapheme-to-phoneme (G2P) models and tools \cite{mortensen2018epitran,hasegawa2020grapheme,novak2016phonetisaurus}, phoneme inventories \cite{phoible}. 
We can relax the requirement of human-validated gold-standard transcripts, and in this paper we obtain the IPA phonetic transcripts by leveraging the LanguageNet G2P models \cite{hasegawa2020grapheme}. 
The LanguageNet G2P models are available for 142 languages, with the phoneme error rates (PERs) ranging from 7\% to 45\%. 
So the main technical aim of this paper is to investigate weakly supervised pretraining with somewhat noisy phonetic transcription. This is in spirit similar to the work in Whisper \cite{whisper}. But instead of using weakly graphemic supervision in Whisper, our work employs weakly phonetic supervision.
We call the approach investigated in this paper: Whistle (\underline{W}eakly p\underline{h}onetic superv\underline{i}sion \underline{st}rategy for multilingua\underline{l} and crosslingual sp\underline{e}ech recognition).

\vspace{+1mm}
We develop Whistle, an approach to data-efficient multilingual and crosslingual speech recognition via weakly phonetic supervision, including the whole pipeline of data processing, model training and testing. Experiments demonstrate the advantages of Whistle for MCL-ASR, in terms of speech recognition for seen languages, crosslingual performance for unseen languages with different amounts of few-shot data, overcoming catastrophic forgetting, and training efficiency.

\RI{Third,} many prior works on multilingual and crosslingual speech recognition were conducted on internal or proprietary datasets such as GlobalPhone \cite{schultz2013globalphone} and IARPA Babel\footnote{\url{https://www.iarpa.gov/index.php/research-programs/babel}}, which are not openly-available.
We find that supervised pretraining with phonetic supervision has been underappreciated so far for MCL-ASR. To promote future research along this direction, we release the code, models and data for the entire pipeline of Whistle at the following URL: \url{https://github.com/thu-spmi/CAT/tree/master/egs/cv-lang10}.

\section{Related work}
\label{sec:related_work}
\subsection{MCL-ASR with phonetic supervision}
Research in multilingual and cross-lingual ASR has long been motivated by phonetics and has used phonetic supervision, e.g., in \cite{schultz1998multilingual,huang2013cross,lu2013cross,li2020universal,zelasko2020sounds, joinAP,xu22b_interspeech}, to name a few.
The major phonetic alphabet in use is the International Phonetic Alphabet (IPA), which includes modified Roman letters and diacritics, by means of which the sounds of all human languages can be represented \cite{fromkin2007introduction}.
So a common practice is to combine the phonetic inventory of all languages to be recognized into a global phoneme set, often based on IPA.
Employing phonetic units is, presumably, the most intuitive way to promote information sharing and learn language-universal representations for MCL-ASR.
Modeling based on phonetic supervision further allows to pursue finer level of information sharing by decomposing phones into a list of phonological articulatory attributes \cite{li2021hierarchical,joinAP,xu22b_interspeech,glocker2023allophant}.

To address the problem of requiring phonetic transcription for phonetic supervision, there have been steady efforts to develop phonetic resources and tools. Epitran provides a 61-language rule-based open-source G2P tool \cite{mortensen2018epitran}; the LanguageNet includes FST (Finite State Transducer) based G2P models in nearly 150 languages \cite{hasegawa2020grapheme}, and PHOIBLE compiles a database of phone inventories for more than 2000 languages and dialects \cite{phoible}. Based on these phonetic resources and tools, there has been continuous studies. Base on Epitran G2P, \cite{li2020universal} first predicts over a shared phone inventory, and then introduces an allophone layer to map into language-specific phonemes. 11 training languages and 2 unseen languages were used. Based on LanuageNet G2P, monolingual, multilingual and (zero-shot) crosslingual CTC models are trained over 13 languages in \cite{zelasko2020sounds}, with the output layer consisting of IPA symbols. Every modifier symbol is treated as a separate token, and so phonetic token error rates (PTERs) are measured. Compared to monolingual models, it reports major PTER improvements across all 13 languages in the multilingual setup, and stark degradation in the crosslingual systems.
The recent studies \cite{li2020universal,zelasko2020sounds} mainly investigate universal phone recognition. There remains an interesting question, as also raised in \cite{zelasko2020sounds}, whether improvements in error rates would also be observed in downstream metrics such as WER. Another related question is which approach of phonetic and graphemic supervision is better for MCL-ASR (RQ-1), since no comparison is conducted in these recent multilingual studies. 

\subsection{MCL-ASR with graphemic supervision}
Graphemic transcription (orthography), as a part of the writing system in a language, does not represent the sounds of a language in a consistent way \cite{fromkin2007introduction}. In many languages, there is a discrepancy between graphemic transcription and phonetic transcription.
With the learning power of deep neural networks, people has begun to build ASR systems with the output layer consisting of graphemic units such as characters \cite{CTCCRF_IC19}, subwords \cite{xiao2018hybrid,zheng2021advancing}, or words \cite{soltau2016neural}, initially for monolingual ASR and 
recently applied to MCL-ASR. 
Using graphemic supervision eliminates the requirement of pronunciation lexicons for different languages and simplifies the pipeline of MCL-ASR. On the other hand, pooling and creating a large set of graphemic tokens from multiple languages brings the label sparsity issue and the resulting MCL-ASR systems tend to be {data-hungry}, and tokenization scheme is an active research question \cite{li2019bytes,meta70}.

Thanks to larger and larger amounts of transcribed speech data and increasingly large neural networks, subword-based supervised pretraining has obtained better and better performance and become a widely adopted strategy in industry to build MCL-ASR systems for increasingly many languages.
For example, the Whisper \cite{whisper} models use the a Byte-Pair Encoding (BPE) text tokenizer and are trained over 680,000 hours cleaned web data by weakly graphemic supervision, capable of recognizing speech from 97 languages.
While achieving impressive performance, recent advances in large MCL-ASR models are presumably an effect of scaling power, and it is hard to argue that the good results are not due to having additional data, nor due to the large neural architecture.
It remains unclear which approach (phonetic supervision or grapheme supervision) is better when evaluated in an equal experimental setting, or if they produce similar results for MCL-ASR. This paper presents our preliminary effort to answer this question (RQ-1).

\vspace{-2mm}
\subsection{MCL-ASR with self-supervision}
\label{sec:related_work_C}
Self-supervised learning methods mainly refer to some recent learning methods based on contrastive learning such as wav2vec 2.0 \cite{baevski2020wav2vec} or masking prediction such as BERT \cite{devlin-etal-2019-bert}, which can still be regarded as unsupervised learning methods from a classical perspective (no data annotation is required). Therefore, the literature often does not strictly distinguish between unsupervised and self-supervised learning methods in terms of terminology, and we can collectively refer to them as unsupervised learning methods.
Self-supervised learning methods such as wav2vec 2.0 \cite{baevski2020wav2vec} have been proposed to learn speech representation in general from multilingual unlabeled speech data.
Based on wav2vec 2.0, XLS-R models \cite{XLS-R} are trained on unlabeled data from 128 languages. 
In the recent Massively Multilingual Speech (MMS) project \cite{MMS}, wav2vec 2.0 based models are pretrained over 1,406 languages, and CTC based multilingual ASR models for 1,107 languages are then finetuned using labeled data for each language. Specifically, a linear layer is added on top of pretrained MMS models which maps to an output vocabulary which is the set of letters in the labeled training data, and is then finetuned with the CTC loss.

As commented in \cite{whisper}, while current unsupervised pretraining has improved the quality of audio encoders, the lack of an equivalently high quality pretrained decoder is a crucial weakness which limits their usefulness. In the following, we provide a closely related comment. We find that current unsupervised pretraining methods in learning audio encoders such as wav2vec 2.0 does not satisfy the so-called \emph{principled unsupervised learning}, since ``the unsupervised objective may be unrelated to the supervised task of interest'' \cite{sutskever2015towards}. In contrast, the GPT based unsupervised pretraining method for natural language processing (NLP) tasks is principled, since the supervised objective is the same as (closely related to) the unsupervised objective but only evaluated on a subset of the sequence in NLP \cite{radford2019language}. 
For ASR tasks, these comments favor supervised pretraining (either grapheme-supervision or phonetic supervision) over the current unsupervised pretraining.
These being said, remarkably, it has been known in various machine learning tasks that supervised and unsupervised training methods are not mutually exclusive and could be jointly used to define semi-supervised learning, e.g., in image classification \cite{song2021empirical}, speech recognition \RI{\cite{bai2022joint,saif2024joint,li2024accent}}, natural language labeling \cite{jrf}, dialog systems \cite{liu2023variational}.
A complete investigation into semi-supervised learning for ASR is outside the scope of this paper.
This paper presents a straightforward empirical comparison between self-supervision and phonetic supervision for MCL-ASR in a common experimental setup (RQ-2).


\begin{figure*}[t]
    \centering
    \scalebox{0.8}{
    \includegraphics[width=1\linewidth]{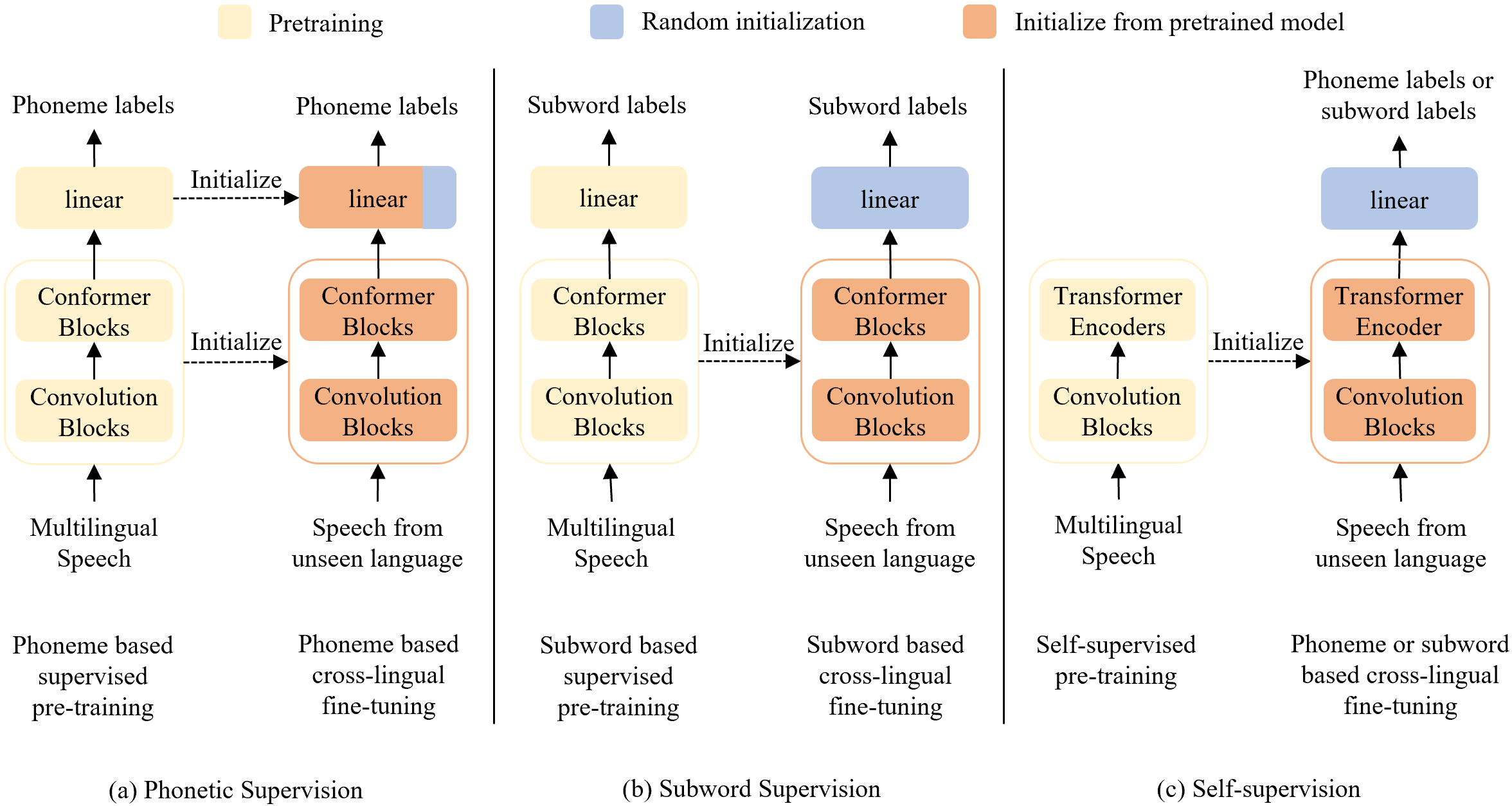}
    }
    \vspace{-2mm}
    \caption{Illustration of the pretraining and finetuning procedures with \RI{(a)} phonetic supervision, \RI{(b)} subword supervision, and \RI{(c)}  self-supervision.}
    \label{fig:overview}
    \vspace{-4mm}
\end{figure*}

\RI{Finally, MCL-ASR with self-supervised, auto-generated phonetic, grapheme or sub-word units based labels is very interesting, which, to our understanding, is closely related to recent progress in unsupervised speech recognition \cite{liu2018completely,baevski2021unsupervised,wang2024unsupervised}, i.e. learning a speech recognizer with only unpaired speech and text. The speech signals are automatically encoded into representative vectors by a pretrained, self-supervised speech encoder such as wav2vec. The representative vectors are then clustered into acoustic tokens, and each speech utterance is represented as a cluster index sequence. It turns out to be very difficult to transcribe the discovered acoustic tokens into phonemes or graphemes in an unsupervised way, though recently there are some progresses, e.g. by GAN \cite{liu2018completely,baevski2021unsupervised} or skipgram and positional unigram matching \cite{wang2024unsupervised}. The dominant way to use pretrained, self-supervised speech encoder is to finetune with phoneme or grapheme labels, as shown in Figure \ref{fig:overview}(c), which is exactly what we evaluate and compare in the experiments. Investigating self-supervised, auto-generated phonetic or graphemic labels, as a related work in unsupervised ASR, is interesting, but out of the scope of this paper.}

\section{Approach}
\label{sec:approach}
In this section, we describe the three main classes of pretraining and finetuning methods for MCL-ASR,  i.e., phoneme-based multilingual supervised pretraining (Section \ref{subsec:phoneme supervision}), subword-based multilingual supervised pretraining (Section \ref{subsec:subword supervision}) and multilingual self-supervised pretraining (Section \ref{subsec:self-supervised pretraining}). 
Figure \ref{fig:overview} shows the differences between the three methods. 
We can see from Figure \ref{fig:overview} that similar neural network architectures can be used for the acoustic encoders in all the three methods, which is good for fair comparison.

The input to the acoustic encoder is usually spectral features, obtained from short-time Fourier transform frame by frame, denoted by $\bm{x}_1,\cdots,\bm{x}_T \triangleq \bm{x}_{1:T}$.
In DNN-based ASR, the acoustic encoder could be viewed as a non-linear feature extractor, which hopefully can be trained to extract high-level features (or say, representations), more discriminative than the raw spectral features.
The output representations from the acoustic encoder are denoted by $\bm{h}_1,\cdots,\bm{h}_T \triangleq \bm{h}_{1:T}$.
A popular neural network architecture for the encoder is Conformer \cite{Conformer}, which consists of convolution blocks followed by Conformer blocks.

Given acoustic observations $\bm{x}_{1:T}$, the task of ASR is to find the most likely labels $y_1,\cdots\,y_L \triangleq y_{1:L}$.
Different units can be used for labeling $y_{1:L}$, depending on what transcription is used for labeling, phonetic or graphemic, as shown in Table \ref{tab:Examples}. Phonemes and subwords are two widely-used labels for MCL-ASR.

In order to promote information sharing between different languages for MCL-ASR, training data from a number of languages, often referred to as seen languages, can be merged to pretrain a multilingual encoder in a supervised fashion, with labels of $y_{1:L}$ given in the form of either phonemes or subwords. Alternatively, the acoustic encoder could be pretrained over unlabeled data by some self-supervised method, such as wav2vec 2.0 \cite{baevski2020wav2vec}, and then be finetuned over labeled data in the form of either phonemes or subwords.

\vspace{-2mm}
\subsection{Phoneme-based multilingual supervised pretraining}
\label{subsec:phoneme supervision}

In this paper, we consider end-to-end ASR models based on the widely used connectionist temporal classification (CTC) method \cite{ctc}.
CTC introduces a blank symbol <b> in addition to the ordinary labels, and further introduces a state sequence $\pi_1,\cdots\,\pi_T \triangleq \pi_{1:T}$, which aids the aligning between $\bm{x}_{1:T}$ and $y_{1:L}$.
Given acoustic sequence $\bm{x}_{1:T}$, at each frame $t$, the possible values that $\pi_t$ can freely take is $V \cup \text{<b>}$, where $V$ denotes the alphabet of labels.
The Conformer based acoustic encoder is used to extract high-level $D$-dimensional representations $\bm{h}_{1:T} = (\bm{h}_1,\cdots\,\bm{h}_T) \in \mathbb{R}^{D \times T}$ from the raw spectral features $\bm{x}_{1:T}$. Then, we can apply a linear layer followed by a softmax activation to calculate the posteriori distribution of $\pi_t$, as follows:
\begin{equation} 
\label{eq:softmax}
\begin{split}
\bm{z}_t &= \bm{W}^T \bm{h}_t \in \mathbb{R}^{|V|+1}\\
P(\pi_t= k |\bm{x}_{1:T} ) &=  \frac{\exp(z_t^k)}{ \sum_{j=1}^{|V|+1} \exp(z_t^j) }, k=1,\cdots,|V|+1
\end{split}
\end{equation}
where $\bm{W} \in \mathbb{R}^{(|V|+1)\times D}$ denotes the weight matrix, and we omit the bias vector in describing the linear layer.
The un-normalized outputs $\bm{z}_t$ are often called logits, and $z^k_t$ denotes the logit corresponding to label $k$. 

In phoneme-based multilingual supervised pretraining investigated in this paper, which is called Whistle, we take the union of the phoneme inventories from the seen languages to be the alphabet of labels $V_{\text{multi}}$.
The $k$-th row vector from the matrix $\bm{W}$, denoted by $\bm{W}(k,:)$, could be viewed as the phoneme embedding for phoneme $k$.
The logit for phoneme $k$ at frame $t$ is actually an inner product between the phoneme embedding and the representation vector, $z_t^k = \bm{W}(k,:)^T \bm{h}_t$.

For recognizing speech from a seen language, the pretrained encoder together with the phoneme embeddings can be directly used without finetuning.
Specifically, we build a weighted finite state transducer (WFST) \cite{mohri2008speech}, obtained by composing the CTC topology, pronunciation lexicon and word-level n-gram language model, and use WFST-based decoding \cite{miao2015eesen,CTCCRF_IC19}.
While requiring pronunciation lexicons (PROLEX), pretraining with phonetic supervision is more advantageous for information sharing between different languages.
In this paper, we relax the requirement of gold-standard human-validated PROLEX and transcripts, by leveraging the LanguageNet G2P models \cite{hasegawa2020grapheme}. The LanguageNet G2P models are available for 142 languages. The phonemization procedure in Whistle is detailed in Section \ref{sec:phonemization}.



For crosslingual speech recognition, denote the phoneme inventory for a new, target language (unseen in pretraining) by $V_{\text{cross}}$.
For recognizing speech from the target language, we can initialize a CTC-based model from the pretrained encoder. The embeddings corresponding to the phonemes in $V_{\text{multi}} \cap V_{\text{cross}}$ are directly copied for initialization.
For those phonemes that are not included in the multilingual phoneme alphabet $V_{\text{multi}}$ but appeared in the target language inventory $V_{\text{cross}}$, we randomly initialize their phoneme embeddings.
The initialized CTC model can then be finetuned over labeled speech from the target language.
In this way, the finetuned encoder and phoneme embeddings can be used to calculate the logits and the posteriori distribution of $\pi_t$ in CTC, and WFST-based decoding can be applied for recognizing speech from the target language.


\vspace{-2mm}
\subsection{Subword-based multilingual supervised pretraining}
\label{subsec:subword supervision}
Multilingual supervised pretraining based on subwords is very similar to that based on phonemes, as described in Section \ref{subsec:phoneme supervision}, which can still base on the CTC method and use WFST-based decoding with word-level n-gram language model.
The major difference is that subword-based multilingual supervised pretraining employs subwords for labeling. Thus, the alphabet of labels $V$ consists of subwords; the lexicon for WFST-based decoing is an orthography lexicon (i.e., words are formed by a sequence of subwords); The row vectors from the matrix $\bm{W}$ could be viewed as embeddings for subwords.
\RI{In crosslingual finetuning of subword-based pretrained models, we employ the common practice to randomly initialize the parameters in the last linear layer. An ablation study is provided in Section \ref{sec:init}, which shows that employing the same initialization scheme as in crosslingual finetuning of phoneme-based pretrained models yields worse performance.}

Converting text into subwords is often referred to tokenization, which is still under investigation and needs a balance between granularity and ASR performance \cite{meta70}.
In this paper, we use Byte Pair Encoding (BPE) based subwords, or say, tokens \cite{sennrich2016neural}. 
BPE introduces a word segmentation algorithm, which initializes the token alphabet with the character alphabet and iteratively merges the most frequent pair of tokens.
In this way, BPE obtains a compact token vocabulary of variable-length subword units.
Notably, the merging of tokens in BPE is based on their frequencies. A straightforward application of BPE may inappropriately favor the merging from high-resource languages; for low-resource languages, tokens may be mostly single characters.
Similar to \cite{conneau2019cross}, sentences are sampled according to a multinomial distribution with probabilities $\left \{ q_{l} \right \}_{l=1...K}$:
\begin{equation}
\label{eq:BPE_sampling}
    q_{l}=\frac{p_{l}^{\beta }}{\sum_{i=1}^{K}p_{i}^{\beta}} \quad with \quad p_{l}=\frac{n_{l}}{\sum_{i=1}^{K}n_{i}},
\end{equation}
where $\beta$ controls the sampling of languages with different frequencies. We use $\beta$ = 0.5 in experiments. $K$ is the number of seen languages in the training data, and $n_l$ denotes the number of sentences for language $l$.
By such data sampling, we can increase the number of tokens associated to low-resource languages and reduce the bias towards high-resource languages.


\vspace{-3mm}
\subsection{Multilingual self-supervised pretraining}
\label{subsec:self-supervised pretraining}
We pretrain a wav2vec 2.0 model \cite{baevski2020wav2vec} on our multilingual pretraining data (just audio data).
The basic architecture of the wav2vec 2.0 model is as follows.
A convolutional feature encoder 
maps raw audio $\bm{x}_{1:T}$ to latent speech features $\bm{z}_{1},\dots,\bm{z}_{T}$, which are then fed to a Transformer 
to output contextual representations $\bm{h}_{1},\dots,\bm{h}_{T}$ \cite{baevski2019vq,devlin-etal-2019-bert}. 
The Transformer architecture is the same as in BERT \cite{NIPS2017_3f5ee243,devlin-etal-2019-bert}. 
During training, a quantization module is employed to discretize the latent features $\bm{z}_{1},\dots,\bm{z}_{T}$ to $\bm{q}_{1},\dots,\bm{q}_{T}$, which represent the targets in the contrastive learning objective. 
The quantization module uses a Gumbel softmax to choose entries from the codebooks and the chosen entries are concatenated to be $\bm{q}_{1},\dots,\bm{q}_{T}$ \cite{jegou2010product,jang2017categorical,baevski2019vq}. 
The wav2vec 2.0 model is trained by solving a contrastive task on masked feature encoder outputs. During training, spans of ten time steps with random starting indices are masked. The objective is to predict the true quantized latent $\bm{q}_{t}$ for masked time-steps within a set of $K = 100$ distractors sampled from other masked time steps.

Basically, the pretrained wav2vec 2.0 model is only an acoustic encoder, consisting of a convolutional feature encoder and a transformer contextual encoder.
In order to recognize speech from any language, we need to introduce a linear layer (parameterized by matrix $\bm{W}$) followed by softmax on top of the encoder output $\bm{h}_1,\cdots,\bm{h}_T$, as shown in Eq. \eqref{eq:softmax}, and perform finetuning over labeled data. The labels could be in the form of either phonemes or subwords.


\section{Experimental setup}
\subsection{Dataset}
We conduct experiments on the CommonVoice dataset \cite{ardila-etal-2020-common} released at September 2022 (v11.0). CommonVoice is a large multilingual speech corpus, with spoken content taken primarily from Wikipedia articles. \RI{It is released under a Creative Commons CC0 license, and has often been used in multilingual speech recognition studies \cite{XLSR,XLS-R,MMS,joinAP,wang2021unispeech,whisper,meta70,meta50}.} We select ten languages for multilingual pretraining experiments: English (en), Spanish (es), French (fr), Italian (it), Kyrgyz (ky), Dutch (nl), Russian (ru), Swedish (sv), Turkish (tr) and Tatar (tt), with a total of 4069.3 hours, which cover rich language families.
We refer to this dataset of 10 languages as CV-Lang10. 
\RI{These ten languages are chosen because they are frequently used to evaluate the performance of multilingual speech recognition systems \cite{XLSR,joinAP,wang2021unispeech}.
We will release the data pre-processing scripts, including text normalization and phonemization, so that people can easily run these scripts themselves to obtain CV-Lang10 from the orignal CommonVoice dataset. Hopefully, in this easy way and based on CC0 license, CV-Lang10 can serve for a common, free experimental setup to facilitate future MCL-ASR research.}
We select Polish (pl) and Indonesian (id) for crosslingual finetuning experiments, which are from two unseen language families. Detailed database descriptions are shown in Table \ref{tab:dataset}. We combine all data from the ten languages to form the training, development, and test sets for multilingual pretraining experiments. For each language, we use its transcripts of training set to separately train a word-level 4-gram language model for WFST-based decoding.

\begin{table}[t]
    \caption{Multilingual and crosslingual data information, including the language code, the language family, the number of IPA phonemes, \RI{characters and BPEs} for each language.}
    \label{tab:dataset}
    \vspace{-2mm}
    \centering
    \scalebox{0.9}{
    \begin{tabular}{p{6mm}<{\centering}|p{5mm}<{\centering}|c|c|p{4mm}<{\centering}|cc}
	\toprule
        & \textbf{Code} & \textbf{Language}& \textbf{Family} & \textbf{IPA} & \RI{Char} & \RI{BPE} \\
        \midrule
        \multirow{10}{*}{Multi.} & en & English & West Germanic & 39& \RI{29} & \RI{3422}   \\
        & es & Spanish & Romance & 32& \RI{89} & \RI{3359}   \\
        & fr & French & Romance & 33& \RI{66} & \RI{3507}   \\
        & it & Italian & Romance & 30& \RI{46} & \RI{3321}  \\
        & ky & Kyrgyz & Turkic & 32 & \RI{37} & \RI{784}   \\
        & nl & Dutch & West Germanic & 39 & \RI{29} & \RI{2304}    \\
        & ru & Russian & East Slavic & 32 & \RI{53} & \RI{974}   \\
        & sv & Swedish & North Germanic & 33& \RI{30} & \RI{2000}   \\
        & tr & Turkish & Turkic & 41& \RI{39} & \RI{1582}   \\
        & tt & Tatar & Turkic & 31& \RI{47} & \RI{773}   \\
        \midrule
        \multirow{2}{*}{Cross.} & pl & Polish & West Slavic & 35 & \RI{34} & \RI{500}   \\
        & id & Indonesian & Austronesian & 35 & \RI{32} & \RI{500}  \\
	\bottomrule
    \end{tabular}
    }
\vspace{-4mm}
\end{table}

\vspace{-3mm}
\subsection{Text normalization, phonemization \RI{and tokenization}}
\label{sec:phonemization}
For text normalization, all punctuation marks are removed, except those that affect pronunciation (such as the apostrophe in English). Certain sentences contain many foreign words are discarded, since G2P converters cannot properly convert them. For reproducible research, details of text normalization and the IDs of deleted sentences for each language will be released in our public repository.

The FST (Finite State Transducer) based G2P toolkit, Phonetisaurus \cite{novak2016phonetisaurus}, is utilized to generate labeling of utterances in IPA phonemes from text transcripts.
The trained FSTs for use with Phonetisaurus can be obtained from LanguageNet \cite{hasegawa2020grapheme}.
Examples of phoneme annotations for each language in CV-Lang10 are shown in Table \ref{tab:Examples}.
By applying Phonetisaurus G2P tool with LanguageNet FSTs, we can also create a PROLEX for each language, which is needed for WFST-based decoding with phoneme-based CTC model.
The \RI{PROLEXs and G2P conversion code} for CV-Lang10 will be released in our public repository.

Remarkably, our phonemization procedure produces weakly phonetic supervision for model training.
The FST-based G2P procedure by LanguageNet and Phonetisaurus is not perfect. As noted in \cite{hasegawa2020grapheme}, PERs ranging from 7\% to 45\%. We only correct a few obvious labeling errors, but the phoneme labels are still somewhat noisy in general.
Additionally, we remove the diacritics and suprasegmentals (like stress and tone) that may be necessary for representing phones, and mainly use base phonemes in our annotation\footnote{From phonetics and phonology \cite{fromkin2007introduction},
while phones represent physical speech sounds (and thus language-independent), phonemes are not physical sounds; they are abstract mental representations of the phonological units of a language, the units used to represent words in our mental lexicon (and thus language dependent).
A particular realization (pronunciation) of a phoneme is called a phone. The collection of phones that are the realizations of the same phonemes are called the allophones of that phoneme.
Phonemes for annotation are thus in a coarser granularity than phones, which may facilitate sharing between languages.
The 12 languages examined in this paper are all non-tonal languages. So we preliminarily sidestep the problem how tones should be incorporated in phoneme-based multilingual models. This is a interesting future work, as previously investigated in \cite{li2020autosegmental}.
}.
While some recent studies pursue universal phone recognition \cite{li2020universal,zelasko2020sounds}, this paper does not aim for phone recognition. 
On the one hand, accurate gold-standard phone labeling is hard to obtain. On the other hand, when we use WFST-based decoding with PROLEXs and aim for reducing word error rates (WERs), the complexity of constructing an allophone layer to transform the language-independent phone distributions to the language-dependent distributions may not be necessary.
Training with weakly phonetic supervision and decoding with PROLEXs, with phonemes serving as an interface between acoustics and text, is found to obtain superior results in MCL-ASR in our experiments.
Presumably, as long as the PROLEXs and the phonetic transcriptions are aligned in some way, weakly phonetic supervision can well drive model learning.

\begin{figure}[!t]
    \centering
    \scalebox{0.8}{
    \includegraphics[width=1\linewidth]{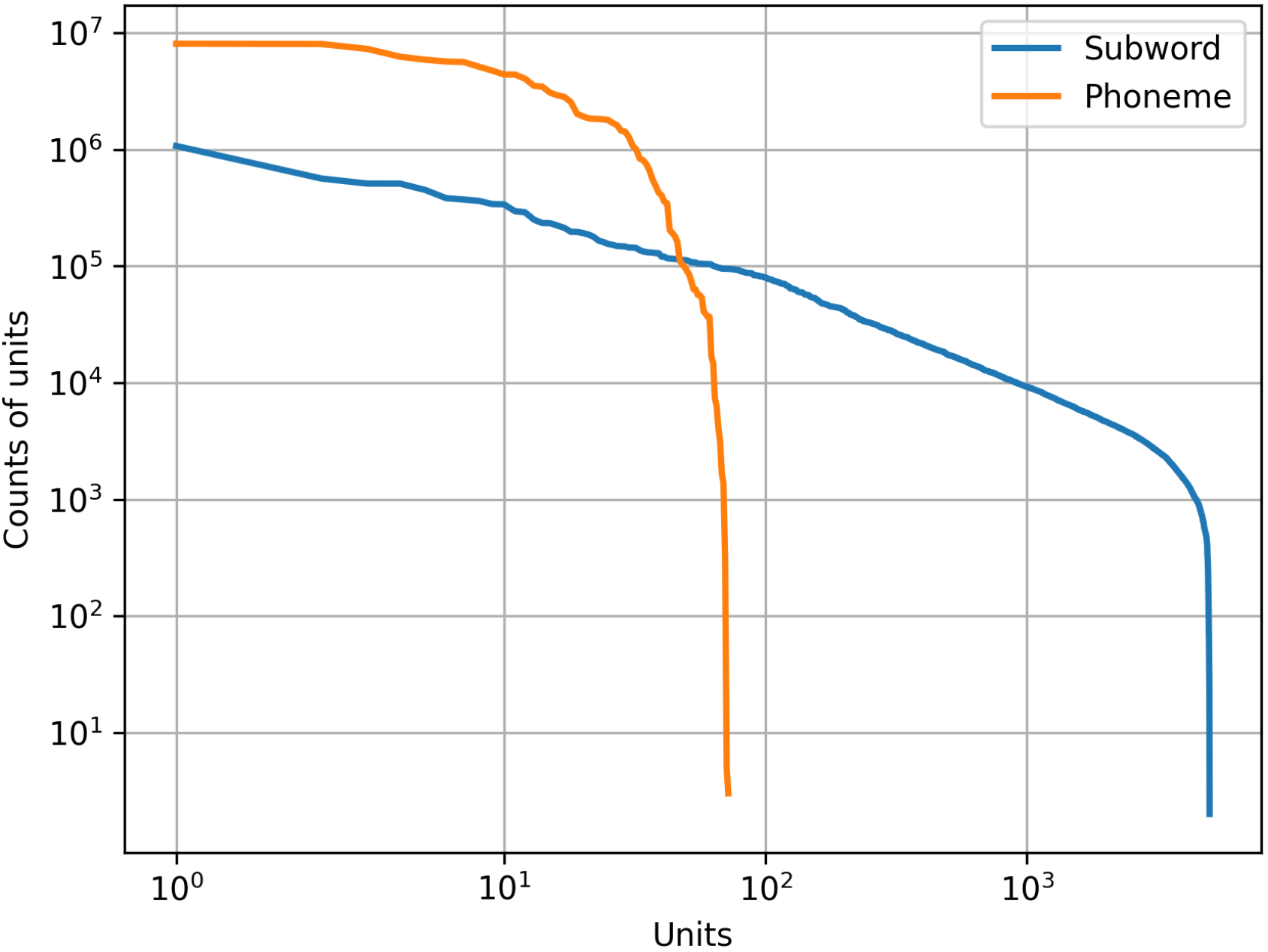}
    }
    \vspace{-3mm}
    \RI{
    \caption{Counts of phoneme and subword units in the CV-Lang10 training set. Note that this is a log-log plot. The distribution of subwords has a sharp peak around a few top subwords and a severe long tail, which shows a more severe data imbalance than the distribution of phonemes.
}
    \label{fig:counting}
    }
    \vspace{-4mm}
\end{figure}

\RI{For phoneme-based models, the multilingual alphabet size of phonemes is 73, which can be naturally determined after phonemization.
For subword-based systems, we use the BPE tokenizer and empirically determine the multilingual BPE vocabulary size to be 4998 after some pilot experiments. There are two additional special tokens <unk> and <s>.
As explained in Section \ref{subsec:subword supervision}, when creating the BPE vocabulary for subword-based supervision, training sentences are sampled according to Eq. \eqref{eq:BPE_sampling}. 
The number of sentences before and after sampling is shown in Table \ref{tab:bpe_sampling} for each language. It can be seen that the numbers of sentences for high-resource languages decrease, while those for low-resource languages increase. 
It can also be seen from Table \ref{tab:dataset} that the resulting multilingual subword vocabulary contains a considerable number of subword units in low-resource languages.
We can see that this sampling strategy alleviates the problem (to some extent) that the frequencies of subwords are severely biased towards high-resource languages.
In summary, the counting statistics for the phoneme and subword units in CV-Lang10 are shown in Figure \ref{fig:counting}.
} 




\begin{table*}[t]
    \caption{Example transcriptions for each language in CV-Lang10}
    \label{tab:Examples}
    \vspace{-2mm}
    \centering
    \begin{tabular}{c|l|l|l}
        \toprule
        \textbf{Code} & \textbf{Text transcript} & \textbf{Transcription with subwords} & \textbf{Transcription with IPA symbols} \\
        \midrule
        en & i know everything about you & i know everything about you & \textscripta \ \textsci \ n o \textupsilon \ \textepsilon \ v \textturnr \ i \textbaro \  \textsci \ \textipa{N} \textschwa \ b a \textupsilon  \ t j u  \\
        es & no lo he visto & no lo he v ist o & n o l o e b i s t o  \\
        fr & vous ne me comprenez pas & vous ne me comp ren ez pas & v y n m k \textopeno \ p \textinvscr \  \textschwa \ n e p a \\
        it & è meglio separarci adesso & è me g lio separ ar ci ad esso & \textepsilon \ m e \textturny \ i o s e p a r a r \texttoptiebar{t\textesh} \  a r s s o \\
        ky & {\fontencoding{OT2}\selectfont менин эч кандай кuн\textbaro\textbaroм жок} & {\fontencoding{OT2}\selectfont мен ин эч кандай кuн \textbaro\textbaro м жок} & m e n i n e \texttoptiebar{t\textesh} k \textscripta \ n d \textscripta \ j k y n ø m \texttoptiebar{d\textyogh} \  o k  \\
        nl & ze is een bekend model & ze is een bek end mod el & z e \textsci \ s e n b \textschwa \ k \textepsilon \ n t m o d \textepsilon \ l \\
        ru & {\fontencoding{OT2}\selectfont база данных обновлена} & {\fontencoding{OT2}\selectfont ба за дан ных об нов лен а} & b a z a d a n \textbari \ x o b n o v l e n a \\
        sv & hörni ta det lugnt & h ör ni ta det lug n t & h œ r n i t \textscripta \ d e t l \textbaru \ \textipa{N} n t \\
        tr & bunlar en büyükleri & bun lar en b üy ük leri & b u n \textltilde \ a \textrtailr \ e n b y j y k l e r i \\
        tt & {\fontencoding{OT2}\selectfont мен\textschwa шулай яш\textschwaп ятабыз} & {\fontencoding{OT2}\selectfont мен \textschwa шулай яш \textschwa п я та быз} & m j e n æ \textesh \ u l a j j a \textesh \ æ p j a t a b \textramshorns \ z \\
        pl & lubię muzykę klasyczną & lu b ię mu zy kę k la sy cz ną & l u a b v i \textepsilon \ m u w z \textbari \ k \textepsilon \ k l a t s \textbari \ \texttoptiebar{\textrtailt \textrtails} n \textopeno \ \textltailn \\
        id & semoga cepat sembuh & sem o ga c ep at sem b uh & s \textepsilon \ m \textupsilon \ \textscriptg \ a \texttoptiebar{t\textesh} \ \textepsilon \ p a t s \textsci \ m b o h \\
        \bottomrule
    \end{tabular}
    \vspace{-2mm}
\end{table*}

\begin{table*}[t]
    \RI{
    \caption{The number of sentences before and after sampling by Eq. (\ref{eq:BPE_sampling}) for each language. The sampled sentences are used to train the BPE tokenizer.}
    \label{tab:bpe_sampling}
    }
    \vspace{-2mm}
    \centering
    \begin{tabular}{l|rrrrrrrrrr|c}
        \toprule
        & \multicolumn{1}{c}{\textbf{en}} & \multicolumn{1}{c}{\textbf{es}} & \multicolumn{1}{c}{\textbf{fr}} & \multicolumn{1}{c}{\textbf{it}} & \multicolumn{1}{c}{\textbf{ky}} & \multicolumn{1}{c}{\textbf{nl}} & \multicolumn{1}{c}{\textbf{ru}} & \multicolumn{1}{c}{\textbf{sv}} & \multicolumn{1}{c}{\textbf{tr}} & \multicolumn{1}{c|}{\textbf{tt}} & \textbf{Total}\\
        \midrule
        Before sampling & 1,583,721 & 274,765 & 607,468 & 188,038 & 26,572 & 61,702 & 106,294 & 28,572 & 62,081 & 20,352 & 2,959,565 \\
        Sampled & 867,689 & 536,136 & 361,104 & 298,887 & 225,392 & 172,169 & 171,133 & 115,987 & 112,677 & 98,391 & 2,959,565 \\
        \bottomrule
    \end{tabular}
    \vspace{-4mm}
\end{table*}

\vspace{-3mm}
\subsection{Model training}
The CAT toolkit \cite{an2020cat} is used for training  CTC \cite{ctc} based ASR models in our experiments.
Three sizes of acoustic encoders are used in our experiments, all based on Conformer \cite{Conformer} networks. 
The small-sized Conformer encoder (S) consists of 14 encoder blocks with dimension 512. We set the self-attention layer to have 4 heads with 36-dimension hidden states, and the feed-forward network (FFN) dimension to 512.
The middle-sized Conformer encoder (M) uses 22 blocks, model dimension 640, FFN dimension 640, attention dimension 160, while the large-sized Conformer encoder (L) uses 22 blocks, model dimension 1024, FFN dimension 1024, attention dimension 224.
For phoneme-based models, the multilingual alphabet size of phonemes is 73.

We train all the models using the Noam optimizer \cite{NIPS2017_3f5ee243} and warm up for the first 10$\%$ of updates. \RI{For different models, we empirically determine a total number of iterations after some pilot experiments, which, to our best, reflects the performance of each model. In order to eliminate the influence of utterance length, the CTC loss is normalized by sequence length, which is the default setting in Pytorch.} We set the dropout rate to 0.1. For data augmentation, we use the spectral augmentation \cite{Spec}. We extract 80-dimension FBank features from audio (resampled to 16KHz) as inputs to the acoustic encoder. A beam size of 16 is used for decoding. 
For model selection, we adopt an early-stop strategy, i.e., when the validation set loss does not decrease for 10 consecutive epochs, we stop training and then averaging the three best-performing checkpoints on the validation set for testing.

By using the fairseq toolkit and following the wav2vec 2.0 base configuration provided by the toolkit\footnote{\url{https://github.com/facebookresearch/fairseq/blob/main/examples/wav2vec/config/pretraining/wav2vec2_base_librispeech.yaml}}, a wav2vec 2.0 model is pretrained over the CV-Lang10 dataset, which is referred to as ``\RI{Wav2vec2} (10 lang)''.
Meanwhile, we also download an existing wav2vec 2.0 base model\footnote{\url{https://dl.fbaipublicfiles.com/fairseq/wav2vec/wav2vec_small.pt}}, which was pretrained over English data and is referred to as ``\RI{Wav2vec2} (En)''.
The two wav2vec 2.0 models have same base architecture, which consists of 12 Transformer blocks, model dimension 768, FFN dimension 3072 and 8 attention heads. \RI{Wav2vec2} (10 lang) uses Adam where the learning rate is warmed up for the first 10$\%$ of updates to a peak of 1e-5.

\begin{table*}[t]
    \caption{\RI{Phoneme error rates (PERs) and Word error rates (WERs)} for phoneme-based monolingual models and multilingual pretrained models on the CV-Lang10 dataset, compared with the subword-based multilingual pretrained model. (S: small, M: middle, L: large)}
    \vspace{-2mm}
    \label{tab:multilingual}
    \centering
    \begin{tabular}{c|c|c|cccccccccc|c}
        \toprule
        \textbf{ID} & \textbf{Model} & \textbf{Size (M)} & \textbf{en} & \textbf{es} & \textbf{fr} & \textbf{it} & \textbf{ky} & \textbf{nl} & \textbf{ru} & \textbf{sv} & \textbf{tr} & \textbf{tt} & \textbf{Avg.} \\
        \midrule
        \midrule
        \multicolumn{3}{c|}{\RI{Number of pretraining hours per language}} & \RI{2227.3} & \RI{382.3} & \RI{823.4} & \RI{271.5} & \RI{32.7} & \RI{70.2} & \RI{149.8} & \RI{29.8} & \RI{61.5} & \RI{20.8} & \RI{4069.3}\\
        \midrule
        \multicolumn{14}{l}{\RI{\emph{PER}}} \\
        \midrule
        O1 & Mono. phoneme & 90 & 7.39 & 2.47 & 4.93 & 2.87 & \textbf{2.23} & 4.60 & \textbf{2.72} & 18.69 & 6.00 & 10.54 & 6.11     \\
        \midrule
        M1 & Multi. phoneme S & 90 & 8.02 & 3.37 & 5.68 & 4.04 & 8.29 & 5.77 & 6.05 & 18.07 & 8.32 & 8.53 & 7.61 \\
        M2 & Multi. phoneme M & 218 & 6.70 & 2.63 & 4.53 & 3.12 & 5.95 & 3.95 & 4.61 & 14.81 & 6.04 & 8.47 & 6.08 \\
        M3 & Multi. phoneme L & 543 & \textbf{5.42} & \textbf{1.96} & \textbf{3.52} & \textbf{2.25} & 4.06 & \textbf{2.64} & 2.97 & \textbf{11.33} & \textbf{4.04} & \textbf{5.97} & \textbf{4.41} \\
        \midrule
        \multicolumn{14}{l}{\RI{\emph{WER}}}\\
        \midrule
        O1 & Mono. phoneme & 90 & 10.59 & 7.91 & 15.58  & 9.26 & 1.03 & 8.84 & 1.62 & 8.37 & 8.46 & 9.75 & 8.14\\
        \midrule
        M4 & Multi. subword & 92 & 12.00 & 9.82 & \textbf{12.40} & 9.98 & 3.29 & 9.67 & 3.31 & 9.95 & 9.11  & 13.56 & 9.30   \\
        \midrule
        M1 & Multi. phoneme S & 90 & 10.76 & 8.68 & 16.01 & 9.98 & 1.02 & 7.32 & 1.59 & 6.14 & 7.63  & 7.30 & 7.64  \\
        M2 & Multi. phoneme M & 218 & 9.83 & 7.82 & 14.94 & 9.04 & \textbf{0.91} & 6.57 & 1.65 & 5.65 & 7.27 & 7.37 & 7.10   \\
        M3 & Multi. phoneme L & 543 & \textbf{8.80} & \textbf{7.02} & 14.02 & \textbf{8.16} & 0.94 & \textbf{6.22}& \textbf{1.46} & \textbf{5.06}& \textbf{7.05}& \textbf{6.92}& \textbf{6.56}\\
        \bottomrule
    \end{tabular}
    \vspace{-6mm}
\end{table*}

\section{Results}
\label{sec:results}
In the following, we introduce the experimental results over CV-Lang10, which serves as a common setup for comparing the three MCL-ASR approaches - 
supervised pretraining with weakly phonetic supervision (Whistle), subword-based supervised pretraining, and wav2vec 2.0 based self-supervised pretraining.
The three approaches are described in Section \ref{subsec:phoneme supervision},
\ref{subsec:subword supervision},
and \ref{subsec:self-supervised pretraining}, respectively.
An MCL-ASR approach is usually evaluated under two tasks. The first is to recognize utterances from seen languages, i.e., the languages that are included in multilingual pretraining. The second is to recognize utterances from unseen languages, i.e., crosslingual speech recognition, which is often performed by finetuning the model obtained from pretraining.

\vspace{-3mm}
\subsection{Multilingual pretraining}
On the CV-Lang10 dataset, 10 phoneme-based monolingual models are trained, each for a single language and with 90M parameters.
Phoneme-based multilingual models (Whistle models) and subword-based multilingual models are trained for comparison.
WFST-based decoding are used for all models.
The PERs and WERs are shown in Table \ref{tab:multilingual}.
The main observations are as follows.

\vspace{+1mm}
\emph{1) Comparing within phoneme-based models}, it can be seen that pooling data from multiple languages and training multilingual models clearly reduces PERs over monolingual models, as shown in prior works \cite{li2020universal,zelasko2020sounds}.
Particularly, a single multilingual model (Mult. phoneme L with 543M parameters) performs significantly better than the 10 monolingual separately-trained models (10 * 90M parameters), on averaged PERs over the 10 seen languages.
Furthermore, we can see that reductions in WERs can be obtained as well, by phoneme-based multilingual pretraining and WFST-based decoding.
Interestingly, in terms of WERs, even the small multilingual model (Mult. phoneme S with 90M parameters) surpasses the  monolingual models.

\vspace{+1mm}
\emph{2) Comparing the phoneme-based and subword-based multilingual models}, it is found that the phoneme-based multilingual model (M1) obtains better WERs than the subword-based multilingual model (M4), with a relative WER reduction of 18\%\footnote{An exception is that for French, the phoneme-based multilingual model does not outperform the subword-based multilingual model in WER, though the PERs are good.
From the statistics of CV-Lang10, we find that the percentage of homophones in the G2P PROLEX of French is the highest (22.5\%). The other large percentages of homophones in the 10 langauges in CV-Lang10 is 9.0\% for English, 5.2\% for Spanish, while others are below 3\%.
Moreover, it is found that some consonants in French words are usually not pronounced, but they may be pronounced when they are spoken in sentences. The WFST-based decoding with a PROLEX may not be good at capturing these regularities. These issues could be alleviated by developing a better method of decoding from phonemes, which will be explored in future.}.
Both models are trained with the same dataset and the same encoder architecture, with close model sizes (around 90M)\footnote{The minor difference in model sizes between phoneme-based model and subword-based model (90M vs 92M) is due to the size of the linear layer because of the different alphabet sizes.}.
\RI{This is a fair comparison to answer RQ-1, representing better \emph{multilingual data-efficiency} of phonetic supervision over graphemic supervision.}
Intuitively, compared to using subwords which mainly serve for text writing, using phonemes as labels is more natural and better for sound classification, since inherently they are more directly related to describing sounds for languages.

\RI{In the following, we provide two points to understand why phonetic supervision obtains better multilingual data-efficiency than graphemic supervision, from the perspectives of data balance and data augmentation respectively.}
First, graphemic supervision suffers from a more severe data imbalance than phoneme supervision, which can be clearly seen from Figure \ref{fig:counting}.
From a machine learning perspective, multi-task learning could be severely affected by data imbalance.
When data are not well balanced in training, an annoying phenomenon, often observed in subword-based systems, is that high resource languages may suffer from interference and low resource languages may be under-trained, which cause performance degradation \cite{li2019bytes,meta70}.
Subword-based systems need special tricks to struggle with data imbalance, such as careful tokenization to appropriately creating the set of tokens \cite{meta70}, human-in-the-loop data mixing in training \cite{li2019bytes}.
In contrast, the superior performances from phoneme-based systems are obtained by training on \emph{natural data mixing} and adopting the classic IPA symbols that have been matured for describing human sounds for a long time.

\begin{figure}[!t]
    \centering
    \scalebox{0.9}{
    \includegraphics[width=1\linewidth]{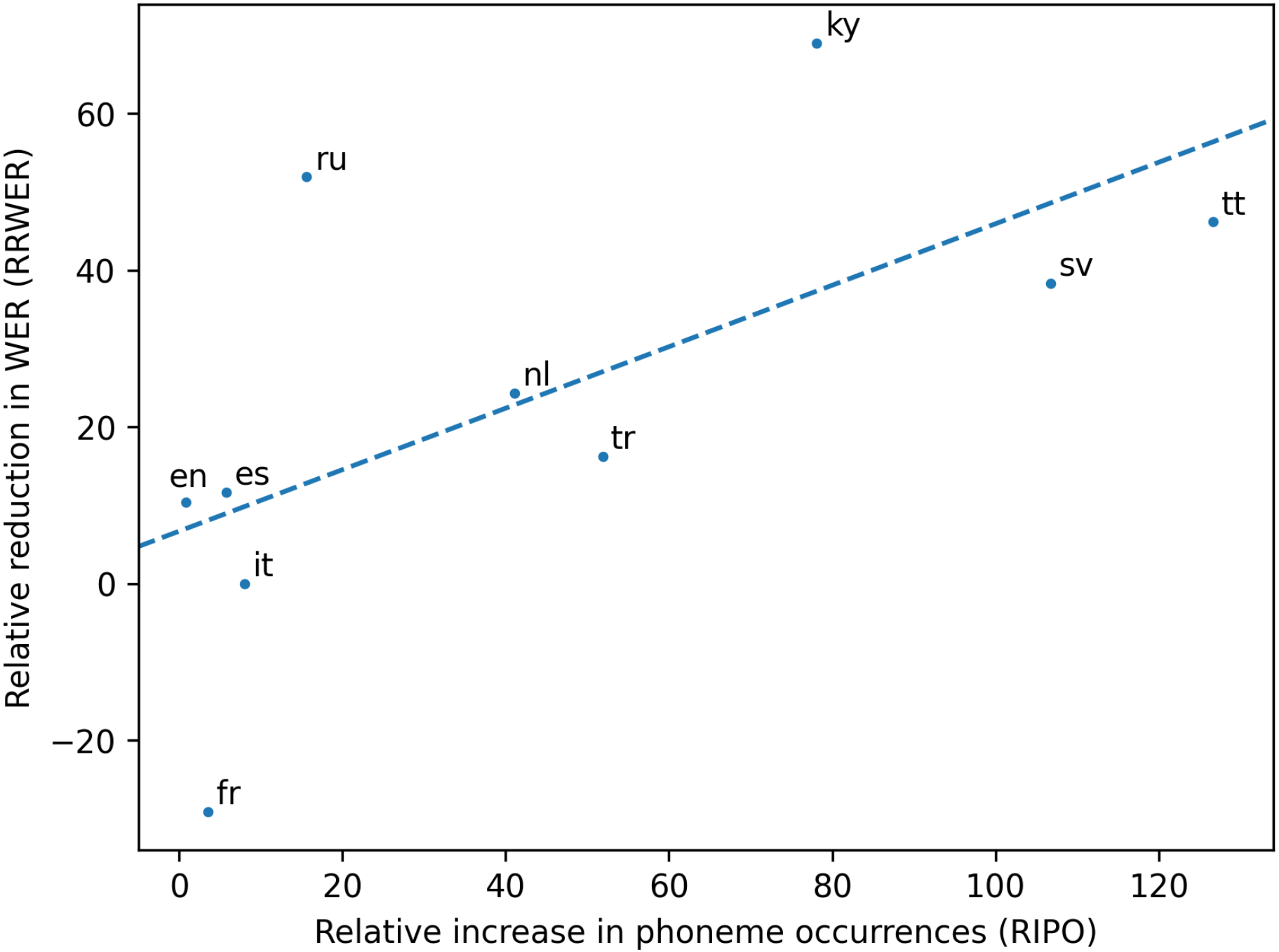}
    }
    \vspace{-2mm}
    \RI{
    \caption{Relative reduction in WER (RRWER) (comparing phoneme pretraining (M1) against subword pretraining (M4) in multilingual speech recognition), as a function of relative increase in phoneme occurrences (RIPO), for the ten languages in CV-Lang10. The figure shows the line of best linear fit: $\text{RRWER} = 0.39 \times \text{RIPO} + 6.6$.}
    \label{fig:relative_WER}
    }
    \vspace{-4mm}
\end{figure}

\RI{Second, phonetic supervision enables a more efficient data sharing than graphemic supervision in multilingual training.
Given a certain amount of training data from a language in multilingual training, the training data for a phoneme in this language is actually augmented in multilingual training, if this phoneme is also occurred in other languages. 
The more sharing in other languages, the more training data is augmented.
This effect can be viewed as \emph{an implicit kind of data augmentation}.
Considering that phonemes are more shared between different languages than subwords, such data augmentation is stronger in phoneme-based multilingual pretraining than in subword-based multilingual pretraining.
Presumably, this explains the better performance of phonetic supervision than graphemic supervision, when both are trained over the same amount of data, i.e. better multilingual data-efficiency.
We can perform a numerical analysis.
For each language, we calculate a base size, i.e., the sum of the number of occurrences of the phonemes in this language in its own data, and an augmented size, i.e., the sum of the number of occurrences of these phonemes by counting over the entire multilingual data.
We then calculate the relative increase in phoneme occurrences (RIPO), by comparing the augmented size against the base size.
Meanwhile, for each seen language, we calculate the relative reduction in WER (RRWER), when phoneme pretraining (M1) is compared against subword pretraining (M4). 
Figure \ref{fig:relative_WER} shows RRWER as a function of RIPO for the ten languages in CV-Lang10.
We can see that RRWER is positively correlated with RIPO.
Low-resource languages obtain more data sharing from multilingual phoneme-based pretraining, enabling them to achieve more relative WER reduction compared to subword-based pretraining.}
 
\vspace{+1mm}
\emph{3) We can see clear scaling properties of phoneme-based models} - PERs and WERs are consistently reduced for both high-resource and low-resource languages, as the model sizes are increased.
Again, remarkably, the performance improvements for different sizes of phoneme-based models are obtained by training on natural data mixing.



\vspace{-3mm}
\subsection{Crosslingual finetuning}
Over the CV-Lang10 dataset, we obtain the phoneme-based supervised pretrained model (M1), which can be further finetuned with either phoneme labels or subword labels for crosslingual speech recognition.
The subword-based supervised pretrained model (M4) is finetuned with subword labels for crosslingual speech recognition. 
The wav2vec 2.0 models, \RI{``Wav2vec2 (10 lang)'' and ``Wav2vec2 (En)''}, can be finetuned with either phoneme labels or subword labels for crosslingual speech recognition.
The four pretrained models used in the crosslingual experiments all have the same model size (around 90M parameters).
On the four pretrained models, we perform full-parameter finetuning, except that for the two wav2vec 2.0 based pretrained models, the convolutional feature encoder are frozen.

To test different multilingual pretrained models for crosslingual speech recognition, we conduct phoneme-based and subword-based crosslingual finetuning on unseen languages.
The training data from an unseen language is divided into three scales to simulate different resource scenarios, while the test and validation data remain unchanged. 


The first unseen language is Polish. 
Polish has 31 phonemes contained in CV-Lang10 and 4 unseen phonemes. The training data is divided into three scales: 1 hour, 10 hours, and full (130 hours). 
\RI{From Table \ref{tab:polish}}, we have the following main observations.
\begin{itemize}
\item In the low-resource scenario with 1-hour Polish training data, phoneme pretraining (PT) followed by phoneme finetuning (FT) performs the best (6.95).
Results with phoneme PT are much better than those with subword PT, which clearly shows the advantage of phonetic supervision in representation learning from  multilingual data (RQ-1).
When comparing phoneme PT and wav2vec 2.0 PT \RI{(M8 vs M6)}, phoneme PT shows obvious superiority (RQ-2).
\item In the scenario with 10-hour Polish training data, the performance with subword PT models begins to improve. When followed by subword FT, both phoneme PT and subword-based PT show equally excellent results (4.83 and 4.89).
\item With the full Polish training data, the wav2vec 2.0 PT models start to perform well, surpassing results with both subword PT and phoneme PT (3.45 < 3.76 < 3.82). This may reflect some benefit of wav2vec 2.0 PT when finetuned with abundant labels, but such top-performing result with wav2vec 2.0 PT is not observed in Indonesian experiments, as shown below.
\end{itemize}

The second unseen language is Indonesian. 
All 35 phonemes of Indonesian are contained in CV-Lang10.
But Indonesian belongs to the Austronesian language family, which are somewhat more different from CV-Lang10, and only 20 hours of training data are available. These make crosslingual finetuning for Indonesian more challenging.  
The training data is divided into three scales: 1 hour, 10 hours, and full (20 hours).

\RI{From Table \ref{tab:indonesian}} for Indonesian, the observations are similar to those for Polish.
In the more challenging scenario with larger linguistic difference and less training data, the advantages of phoneme PT followed by phoneme FT are more obvious, across all the three scales of data settings.
It seems that when training data are more limited, the better results can be obtained by phoneme supervision, compared to subword supervision and self-supervision. 
When the amount of crosslingual training data increases, the performance gaps between phoneme supervision, subword supervision and self-supervision may diminish. Presumably, the finetuning with abundant data behaves like end-to-end monolingual training and the effect of different PT methods may become weak.
\RI{In summary, we find that compared to both graphemic supervision and self-supervision, phonetic supervision excels in crosslingual data-efficiency. The efficiency advantage is more significant when the data are more limited and may diminish when the finetuning data are abundant.}




\begin{table*}[t]
    \caption{Error rates for \RI{phoneme-based and subword-based} crosslingual finetuning (FT) on \textbf{Polish}. The pretraining (PT) dataset is CV-Lang10. For phoneme FT, we report PERs for beam search without PROLEX and WERs for WFST decoding, respectively. For subword FT, we report WERs for beam search without LM and WFST decoding with LM, respectively.}
    \vspace{-2mm}
    \label{tab:polish}
    \centering
    \begin{tabular}{c|l|p{7mm}<{\centering}p{7mm}<{\centering}|p{7mm}<{\centering}p{7mm}<{\centering}|p{7mm}<{\centering}p{7mm}<{\centering}|p{7mm}<{\centering}p{7mm}<{\centering}|p{7mm}<{\centering}p{7mm}<{\centering}|p{7mm}<{\centering}p{7mm}<{\centering}}
        \toprule
        \multirow{4}{*}{\textbf{ID}} & \multirow{4}{*}{\textbf{Pretrained Model}} & \multicolumn{6}{c|}{\textbf{Phoneme FT (PER/WER)}} & \multicolumn{6}{c}{\textbf{Subword FT (WER)}} \\
        &  & \multicolumn{2}{c|}{\textbf{1 hour}} & \multicolumn{2}{c|}{\textbf{10 hour}} & \multicolumn{2}{c|}{\textbf{130 hour}}  & \multicolumn{2}{c|}{\textbf{1 hour}} & \multicolumn{2}{c|}{\textbf{10 hour}} & \multicolumn{2}{c}{\textbf{130 hour}}\\
        &  & \multirow{2}{*}{\textbf{PER}} & \multirow{2}{*}{\textbf{WER}} & \multirow{2}{*}{\textbf{PER}} & \multirow{2}{*}{\textbf{WER}} & \multirow{2}{*}{\textbf{PER}} & \multirow{2}{*}{\textbf{WER}} & \textbf{w/o} & \textbf{~w} & \textbf{w/o} & \textbf{~w}&\textbf{w/o} & \textbf{~w}\\
        & & & & & & & &\textbf{LM}&\textbf{LM}&\textbf{LM}&\textbf{LM}&\textbf{LM}&\textbf{LM} \\
        \midrule
	O2 & Mono. & 86.01 & 99.98 & 30.38 & 13.86 & 2.82 & 4.97 & 98.41 & 98.38 & 90.98 & 59.43 & 19.38 & 7.12\\
        \midrule
        M5 & Wav2vec2 (En) & 25.76 & 11.09 & 16.64& ~6.75 & 5.80 & 4.57 & 100 & 100 & 45.64 & 7.08 & 8.53 & 3.85 \\
	M6 & Wav2vec2 (10 lang) & 21.10& ~7.94& 12.65& ~5.65& 6.08 & 4.44 & 99.97& 100 & 36.93& 5.71& 7.49& \textbf{3.45}\\
        M7 & M4 (subword PT)  &-&-&-&-&-&-& 70.13 & 9.16 & 31.90 & 4.89 & \textbf{5.44} & 3.76 \\
	M8 & M1 (phoneme PT) & \textbf{17.96}&  ~\textbf{6.95} & \textbf{10.47} &  ~\textbf{5.27}& \textbf{1.97} & \textbf{4.30} & \textbf{69.50}& \textbf{8.63}& \textbf{31.89}& \textbf{4.83} & 5.84 & 3.82 \\
	\bottomrule
    \end{tabular}
    \vspace{-2mm}
\end{table*}

\begin{table*}[t]
    \caption{Error rates for \RI{phoneme-based and subword-based} crosslingual finetuning (FT) on \textbf{Indonesian}. The pretraining (PT) dataset is CV-Lang10.}
    \vspace{-2mm}
    \label{tab:indonesian}
    \centering
    \begin{tabular}{c|l|p{7mm}<{\centering}p{7mm}<{\centering}|p{7mm}<{\centering}p{7mm}<{\centering}|p{7mm}<{\centering}p{7mm}<{\centering}|p{7mm}<{\centering}p{7mm}<{\centering}|p{7mm}<{\centering}p{7mm}<{\centering}|p{7mm}<{\centering}p{7mm}<{\centering}}
        \toprule
        \multirow{4}{*}{\textbf{ID}} & \multirow{4}{*}{\textbf{Pretrained Model}} & \multicolumn{6}{c|}{\textbf{Phoneme FT (PER/WER)}} & \multicolumn{6}{c}{\textbf{Subword FT (WER)}} \\
        &  & \multicolumn{2}{c|}{\textbf{1 hour}} & \multicolumn{2}{c|}{\textbf{10 hour}} & \multicolumn{2}{c|}{\textbf{20 hour}}  & \multicolumn{2}{c|}{\textbf{1 hour}} & \multicolumn{2}{c|}{\textbf{10 hour}} & \multicolumn{2}{c}{\textbf{20 hour}}\\
        &  & \multirow{2}{*}{\textbf{PER}} & \multirow{2}{*}{\textbf{WER}} & \multirow{2}{*}{\textbf{PER}} & \multirow{2}{*}{\textbf{WER}} & \multirow{2}{*}{\textbf{PER}} & \multirow{2}{*}{\textbf{WER}} & \textbf{w/o} & \textbf{~w} & \textbf{w/o} & \textbf{~w}&\textbf{w/o} & \textbf{~w}\\
        & & & & & & & &\textbf{LM}&\textbf{LM}&\textbf{LM}&\textbf{LM}&\textbf{LM}&\textbf{LM} \\
        \midrule
	O3 & Mono. & 96.52 & 100 & 27.30 & 7.71 & 5.74 & 3.28 & 96.62&96.42 &69.57 &49.67  &31.96  &10.85\\
        \midrule
        M9 & Wav2vec2 (En) & 31.30 & 6.73 & 10.89 & 3.31 & 6.84 & 2.83 &100  &100  &19.98 &5.28  &\textbf{11.68}  &3.59\\
	M10 & Wav2vec2 (10 lang) & 24.91 & 3.75 & 10.32 & 2.79 & 6.30 & 2.47 &99.64& 99.97&19.08&  4.52& 12.01&  3.15\\
        M11 & M4 (subword PT) &-&-&-&-&-&-& \textbf{64.00} &\textbf{23.56}  &19.41 &3.91  &13.15  &3.07 \\
	M12 & M1 (phoneme PT) & \textbf{21.64}& \textbf{3.27}& \textbf{7.90}& \textbf{2.54}& \textbf{4.79}& \textbf{2.43} &67.71  &24.57  &\textbf{18.21} &\textbf{3.59}  &12.48  &\textbf{2.92}\\
	\bottomrule
    \end{tabular}
    \vspace{-2mm}
\end{table*}

\section{Ablation study}
\subsection{Analysis of embeddings}
\begin{figure*}[!t]
    \begin{center}
    \scalebox{0.8}{
    \begin{minipage}{0.5\linewidth}
        \vspace{3pt}
        \centerline{\includegraphics[width=\textwidth]{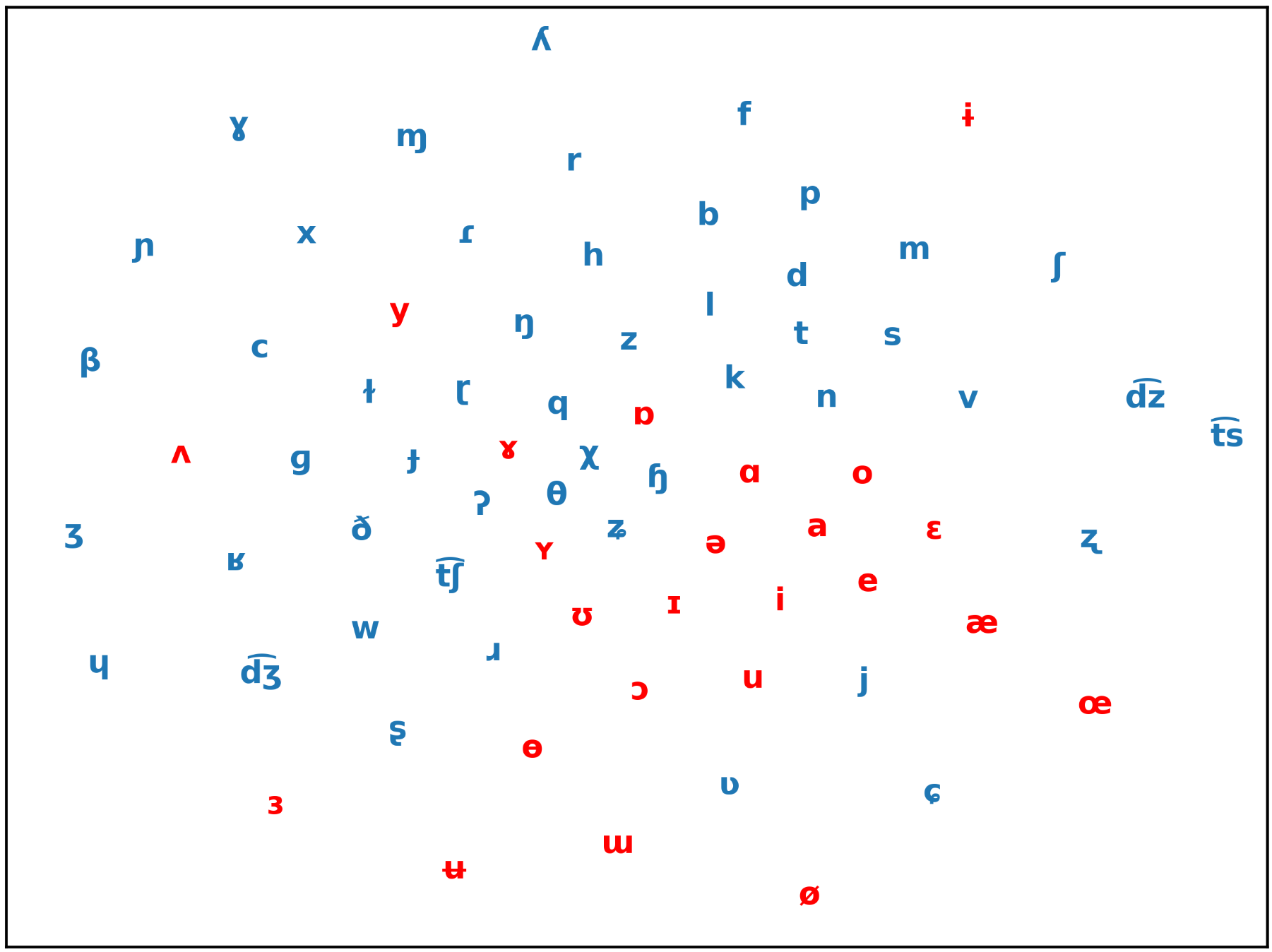}}
        \centerline{\scriptsize (a)}
    \end{minipage}
    \begin{minipage}{0.5\linewidth}
        \vspace{3pt}
        \centerline{\includegraphics[width=\textwidth]{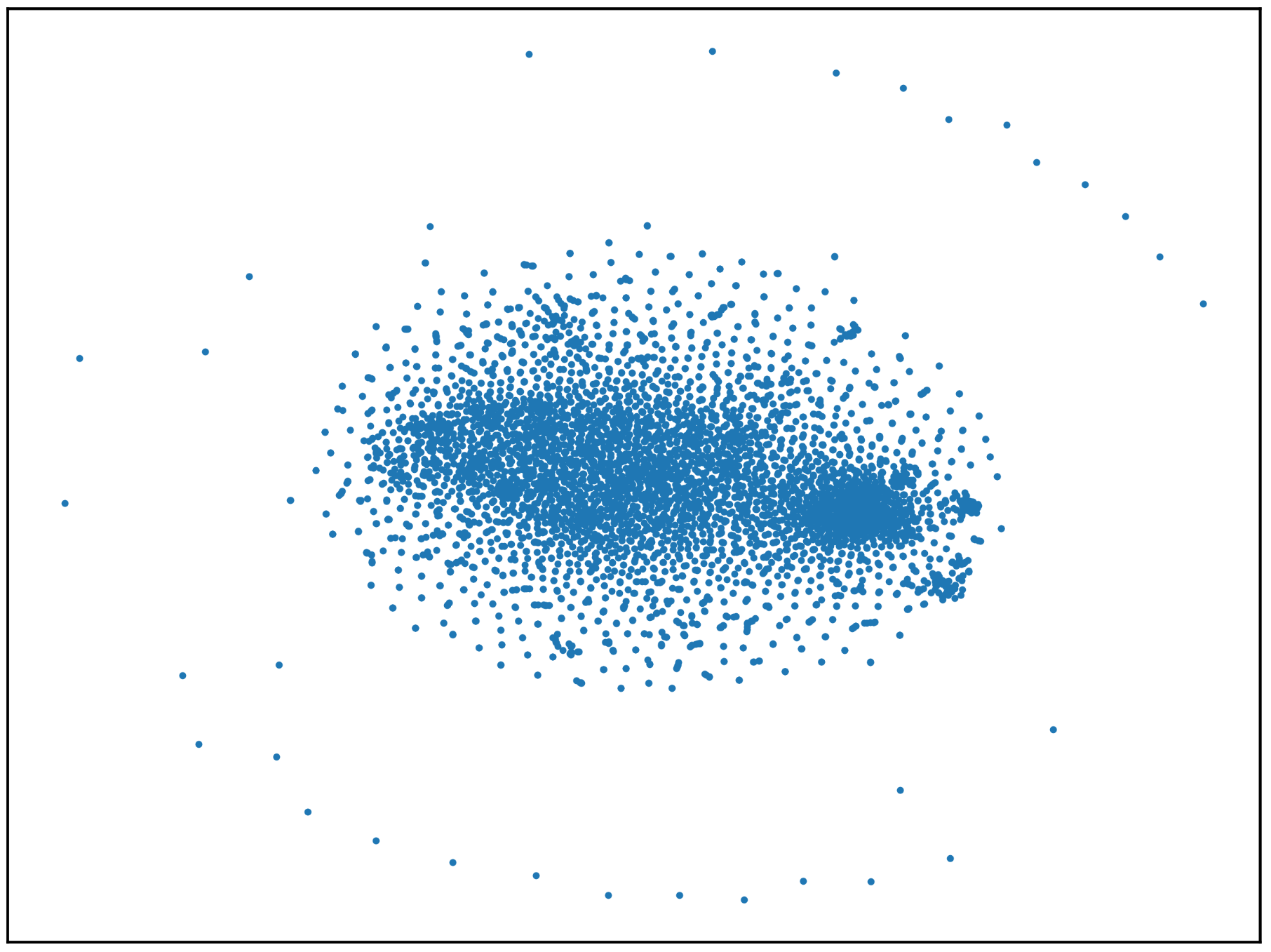}}
        \centerline{\scriptsize (b)}
    \end{minipage}
    }
    \end{center}
    \vspace{-2mm}
    \caption{Visualization of embeddings by t-SNE. (a) Phoneme embeddings from M1, (b) Subword embeddings from M4. In (a), blue indicate the consonants and red indicate the vowels.}
    \label{fig:embeding}
    \vspace{-4mm}
\end{figure*}

To gain intuitive understanding of the multilingual models trained under phonetic supervision and graphemic supervision, we apply t-SNE \cite{van2008visualizing} to draw the 512-dimensional embeddings on a 2-dimensional map. Figure \ref{fig:embeding}(a) and (b) show the maps of the 73 phoneme embeddings and the 4998 subword embeddings, obtained from the phoneme-based model M1 and subword-based model M4, respectively.
By comparing the two figures, it can be easily seen that the phoneme embeddings are more evenly dispersed in the high-dimensional space.
In contrast, subword embedings are densely crowded in the center and become sparser as they move outward.
This indicates that the representation learning in the subword-based model is not so balanced as in the phoneme-based model.
Presumably, this is due to the severe data imbalance in subword supervision.
Furthermore, it can be noticed that most of the vowels embeddings cluster in the bottom right area of Figure \ref{fig:embeding}(a). Certain consonant phonemes, like approximants ('\textturnr', '\textscriptv' and 'j'), 
also appear in this region, since approximants fall between fricatives and vowels.
This reflects that the phoneme-based model not only learns the differences between phonemes, but also captures some phonetic similarities between phonemes.


\begin{table*}[t]
    \caption{Test of catastrophic forgetting for the multilingual models, pretrained over CV-Lang10 and finetuned on 10 minutes of a new language (Polish). WARD denotes word accuracy relative degradation of the averaged WER over the ten old languages in CV-Lang10.}
    \label{tab:Catastrophic forgetting}
    \vspace{-2mm}
    \centering
    \begin{tabular}{c|c|cccccccccc|c|c}
        \toprule
        \textbf{Model} &\hfil \textbf{pl} &\hfil \textbf{en} &\hfil \textbf{fr} & \textbf{es} & \textbf{it} & \textbf{ru} & \textbf{nl} & \textbf{tr} & \textbf{ky} & \textbf{sv} & \textbf{tt} & \textbf{Avg.} & \textbf{WARD}\\
        \midrule
        M1 + 10min phoneme FT  & 11.0 & 68.5 & 69.3 & 57.1 & 50.3 & 48.3 & 60.9 & 31.8 & 58.4 & 42.3 & 33.0 & 52.0 &  48   \\
        M4 + 10min subword FT & 93.2 & 92.2 & 95.0 & 92.5 & 92.5 & 262.5 & 103.6 & 241.5 & 125.9 & 180.5 & 254.4 & 154.1 & 160 \\
        \bottomrule
    \end{tabular}
    \vspace{-4mm}
\end{table*}

\begin{table}[t]
    \caption{Training efficiency of phoneme-based and subword-based pretraining (PT) and finetuning (FT).}
    \label{tab:Training efficiency}
    \vspace{-2mm}
    \centering
    \begin{tabular}{c|c|c}
        \toprule
        \textbf{Model}&\hfil \textbf{Batch size} &\hfil \textbf{Epochs for converging} \\
        \midrule
        M1 & 640 & 63    \\
        M1 + pl subword FT & 320 & 195     \\
        \midrule
        M4 & 640 & 83     \\
        M4 + pl subword FT & 320 & 223     \\
        \bottomrule
    \end{tabular}
    \vspace{-4mm}
\end{table}

\vspace{-3mm}
\subsection{Test of catastrophic forgetting}
\label{sec:catastrophic}


In previous sections, we show the advantage of multilingual pretrained models by phoneme supervision over those by subword supervision for recognizing seen and unseen languages.
We see that after a pretrained multilingual model is finetuned over data from a new language, the finetuned multilingual model can recognize speech from the new language.
Then, to what degree the performance of the finetuned multilingual model on previous seen languages would be affected?
This is an interesting question for continual pretraining of multilingual models to support more new languages, a question related to catastrophic forgetting of neural network based models \cite{mccloskey1989catastrophic}.
A complete investigation into continual pretraining of multilingual models is outside the scope of this paper. 
Here we present a preliminary examination of the two approaches, phoneme or subword-based multilingual models, in overcoming catastrophic forgetting.

\vspace{+2mm}
The phoneme-based multilingual model M1 and the subword-based multilingual model M4, both pretrained over CV-Lang10 and with 90M parameters, are finetuned separately on 10 minutes of a new language (Polish). The finetuned models are then tested not only on Polish, but also on the ten languages in CV-Lang10.
The results are shown in Table \ref{tab:Catastrophic forgetting}.
Phoneme PT followed by 10 minutes of phoneme FT obtains WER of 11.0$\%$ on Polish, while showing a word accuracy relative degradation (WARD) of 48$\%$\footnote{$(52.0-7.61)/(100-7.61)=48\%$} for the averaged WER over the ten old languages in CV-Lang10.
In contrast, subword PT followed by 10 minutes of subword FT yields much worse result for Polish, and actually breaks down in recognizing the ten old languages, totally losing their multilingual recognition ability after finetuning on 10 minutes of a new language.
This suggests that phoneme PT and FT are more robust in overcoming catastrophic forgetting, presumably because the learned representations are stabler and more universal than those learned by subword PT and FT.
Meanwhile, it shows that continual pretraining of multilingual models is a non-trivial problem, which deserves more investigations.


\vspace{-3mm}
\subsection{Training efficiency}
Besides the performance advantage of phoneme-based supervision over subword-based supervision, we find that phoneme-based models tend to be more training efficient, i.e., they can converge with fewer optimization steps.
Table \ref{tab:Training efficiency} shows the training epochs when different models converge.
Under equal batch sizes, phoneme PT takes less training epochs than subword PT, with 24$\%$ reduction.
When crosslingual subword FT is performed on Polish full data, finetuning the phoneme PT model achieves 12$\%$ reduction in finetuning epochs relative to finetuning the subword PT model. 
This finding again reveals that phoneme labels can provide more efficient supervision for sound classification than subword labels. It takes a longer, less efficient path for neural networks to learn sound classification from subword supervision.

\vspace{-6mm}
\RI{
\subsection{Last linear layer initialization in finetuning from subword-based pretrained models}
\label{sec:init}
As introduced in Section \ref{sec:approach}, the row vectors from the weight matrix of the last linear layer can be viewed as embedding vectors for output units, which can be phonemes or subwords.
Denote the union of the unit inventories from the seen languages in pretraining by $V_{\text{multi}}$, and the unit inventory for an unseen language in finetuning by $V_{\text{cross}}$, respectively.
In crosslingual finetuning, the parameters corresponding to seen units in $V_{\text{multi}} \cap V_{\text{cross}}$ can be initialized by directly copying from the pretrained model and those parameters for unseen units are randomly initialized.
This is exactly what we do in finetuning from phoneme-based pretrained models in our experiments.}

\vspace{+8mm}

\RI{
For finetuning from subword-based pretrained models, this similar initialization scheme can be used as well, i.e., copying for seen subwords and random initialization for the remaining subwords.
As suggested by a referee, we compare this scheme to the completely random initialization scheme.
The comparison results are shown in Table \ref{tab:subword FT init}. 
It can be seen that the initialization scheme with copying for seen units\footnote{\RI{Among the 500 BPE subwords for Polish, there are 202 seen subwords in the pretraining data (i.e. CV-Lang10); and for the 500 BPE subwords for Indonesian, there are 234 seen subwords in CV-Lang10.}} performs worse than random initialization for finetuning from subword-based pretrained models.
Presumably, this is because when the same subword occurs in different languages, the sounds pronounced can often vary significantly\footnote{\RI{For example, the subword ``nat'' in the English word ``nature'' is pronounced as [ne\textsci \texttoptiebar{t\textesh}], while the subword ``nat'' in the Polish word ``natomiast'' is pronounced as [nat].}}. The model may be confused in learning. Initialization with copying from seen units does not improve the learning in this case.
Therefore, random initialization is the common practice in finetuning from subword-based pretrained models, which is exactly what we do in our experiments and reflects its best performance.
The comparison of finetuning from phoneme-based and subword based pretrained models in our experiments as shown in Table \ref{tab:polish} and Table \ref{tab:indonesian} is sound.
This further strengthens our basic observation: the essential role of subwords is primarily for the writing of a language, rather than for describing and distinguishing sounds, for which phonemes are defined.
}

\begin{table*}[t]
    \RI{
    \caption{Word error rates for crosslingual finetuning from the subword-based pretrained model (M4) with different initialization schemes for the Last linear layer. Note that the results by random initialized are taken from Table \ref{tab:polish} and Table \ref{tab:indonesian}, which are shown here for clear side-by-side comparison.}
    \label{tab:subword FT init}
    }
    \vspace{-2mm}
    \centering
    \begin{tabular}{l|p{7mm}<{\centering}p{7mm}<{\centering}|p{7mm}<{\centering}p{7mm}<{\centering}|p{7mm}<{\centering}p{7mm}<{\centering}|p{7mm}<{\centering}p{7mm}<{\centering}|p{7mm}<{\centering}p{7mm}<{\centering}|p{7mm}<{\centering}p{7mm}<{\centering}}
        \toprule
        \multirow{4}{*}{\textbf{Exp}} & \multicolumn{6}{c|}{\textbf{Polish}} & \multicolumn{6}{c}{\textbf{Indonesian}} \\
        & \multicolumn{2}{c|}{\textbf{1 hour}} & \multicolumn{2}{c|}{\textbf{10 hour}} & \multicolumn{2}{c|}{\textbf{130 hour}}  & \multicolumn{2}{c|}{\textbf{1 hour}} & \multicolumn{2}{c|}{\textbf{10 hour}} & \multicolumn{2}{c}{\textbf{20 hour}}\\
        & \textbf{w/o} & \textbf{~w} & \textbf{w/o} & \textbf{~w}&\textbf{w/o} & \textbf{~w} & \textbf{w/o} & \textbf{~w} & \textbf{w/o} & \textbf{~w}&\textbf{w/o} & \textbf{~w}\\
        &\textbf{LM}&\textbf{LM}&\textbf{LM}&\textbf{LM}&\textbf{LM}&\textbf{LM} &\textbf{LM}&\textbf{LM}&\textbf{LM}&\textbf{LM}&\textbf{LM}&\textbf{LM} \\
        \midrule
        Random initialization & \textbf{70.13} & \textbf{9.16} & \textbf{31.90} & \textbf{4.89} & \textbf{5.44} & \textbf{3.76} & 64.00 & \textbf{23.56} & \textbf{19.41} & \textbf{3.91} & \textbf{13.15}  & \textbf{3.07} \\
	  Copying for seen units & 71.39 & 12.73 & 32.11 & 5.39 & 6.54 & 4.17 & \textbf{57.08} & 24.84 & 19.72 & 4.99 & 13.40 & 3.40 \\
	\bottomrule
    \end{tabular}
    \vspace{-4mm}
\end{table*}

\section{Conclusions and future work}
This paper starts from examining the pros and cons of the three main approaches for MCL-ASR - supervised pretraining with phonetic transcription or graphemic transcription, and self-supervised pretraining.
We find that pretraining with phonetic supervision has been underappreciated so far for MCL-ASR, while conceptually it is more advantageous for information sharing between different languages.
This paper explores the approach of pretraining with weakly phonetic supervision towards data-efficient MCL-ASR, which is called Whistle.
We relax the requirement of gold-standard human-validated phonetic transcripts, and obtain IPA based transcripts by leveraging Phonetisaurus (an FST based G2P toolkit) with LanguageNet G2P FSTs.
We construct a common experimental setup based on the CommonVoice dataset, called CV-Lang10, with 10 seen languages and 2 unseen languages (Polish and Indonesian). 
A set of experiments are conducted on CV-Lang10 to compare, as fair as possible, the three approaches under the common setup for MCL-ASR. 
Training with weakly phonetic supervision (though somewhat noisy) and decoding with PROLEXs, with phonemes serving as an interface between acoustics and text, is found to obtain superior results in MCL-ASR in our experiments, in terms of speech recognition for seen languages, crosslingual performance for unseen languages with different amounts of few-shot data, overcoming catastrophic forgetting, and training efficiency.
Moreover, phoneme-based models naturally overcome language imbalance and can be efficiently trained on natural data mixing, while subword-based models need careful tokenization and data mixing in training.
When training data is more limited, phoneme supervision can achieve better results compared to subword supervision and self-supervision, thereby providing higher data-efficiency.

This work demonstrates some advantages of weakly phonetic supervision towards data-efficient MCL-ASR. There are interesting directions for future work. 
First, we preliminarily sidestep the problem how tones should be incorporated in pretraining multilingual phoneme-based models, since the 12 languages examined in this paper are all non-tonal languages. There have been some effort towards addressing this problem \cite{li2020autosegmental}.
Second, this work mainly uses WFST based decoding with PROLEXs. Better methods of decoding from phonemes could be explored in future, such as based on sequence-to-sequence models \cite{sutskever2014sequence}.
Third, scaling the approach of Whistle with more languages and more data is expected to achieve increasingly better MCL-ASR performance. Meanwhile, it is worthwhile to investigate how to incrementally learn from new languages with a non-stationary stream
Continual learning methods such as based on prompt pool 
\cite{van2022three,liu2023prompt} could be incorporated into MCL-ASR.

\bibliographystyle{IEEEtran}
\bibliography{references.bib}
 




\vspace{-3cm}

\begin{IEEEbiography}[{\includegraphics[width=1in,height=1.25in,clip,keepaspectratio]{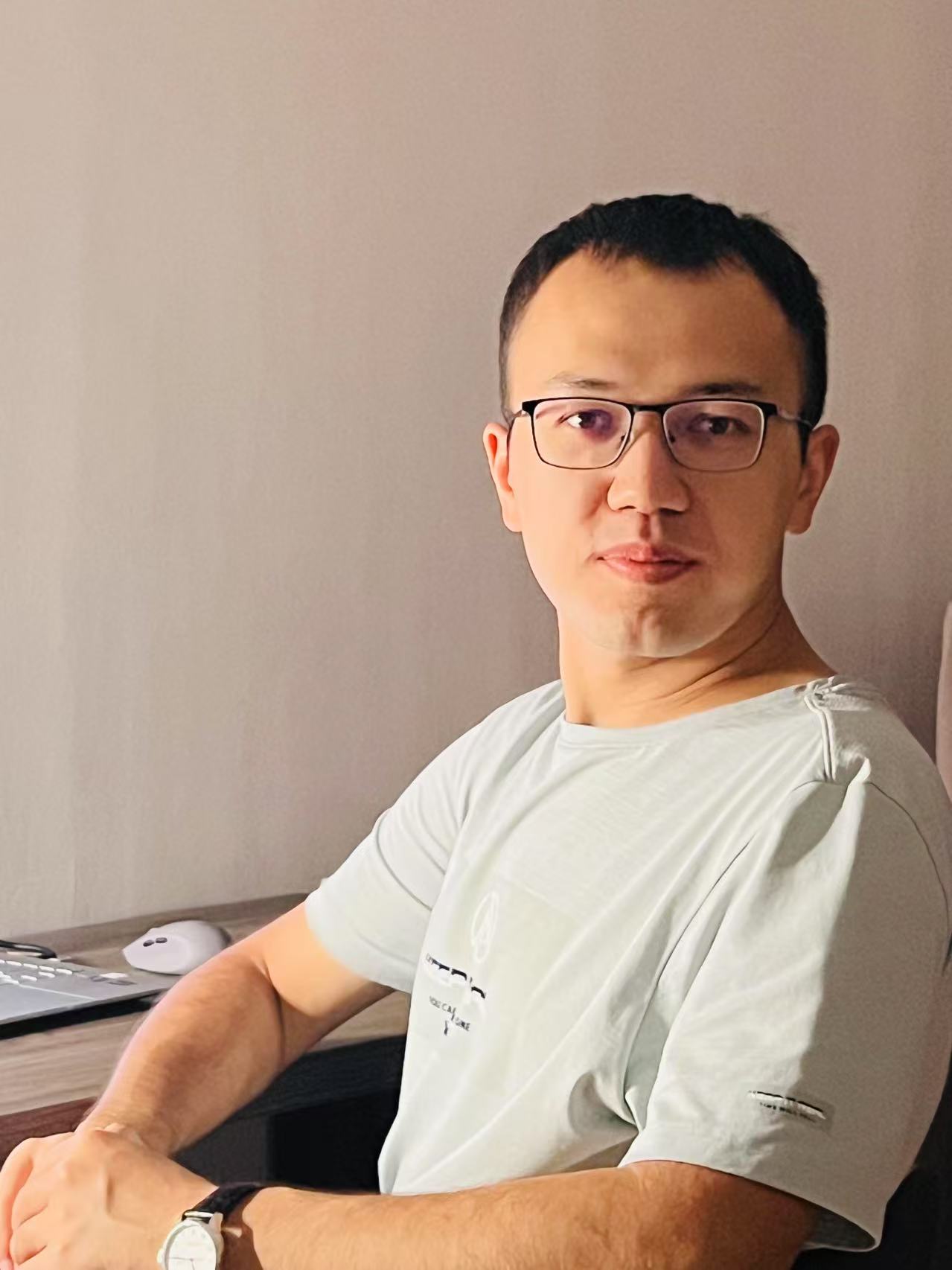}}]{Saierdaer Yusuyin}
received the B.E. degree in software engineering from Northwestern Polytechnical University, Xi'an,
China, in 2019. Since 2019, he has been working toward the Ph.D. degree with the School of Computer Science and Technology, Xinjiang University, under the supervision of Hao Huang and Zhijian Ou (during internship at THU-SPMI since January 2023). His research interests include multiliangual speech recognition and semi-supervised learning theory. 
\end{IEEEbiography}

\vspace{-3cm}

\begin{IEEEbiography}[{\includegraphics[width=1in,height=1.25in,clip,keepaspectratio]{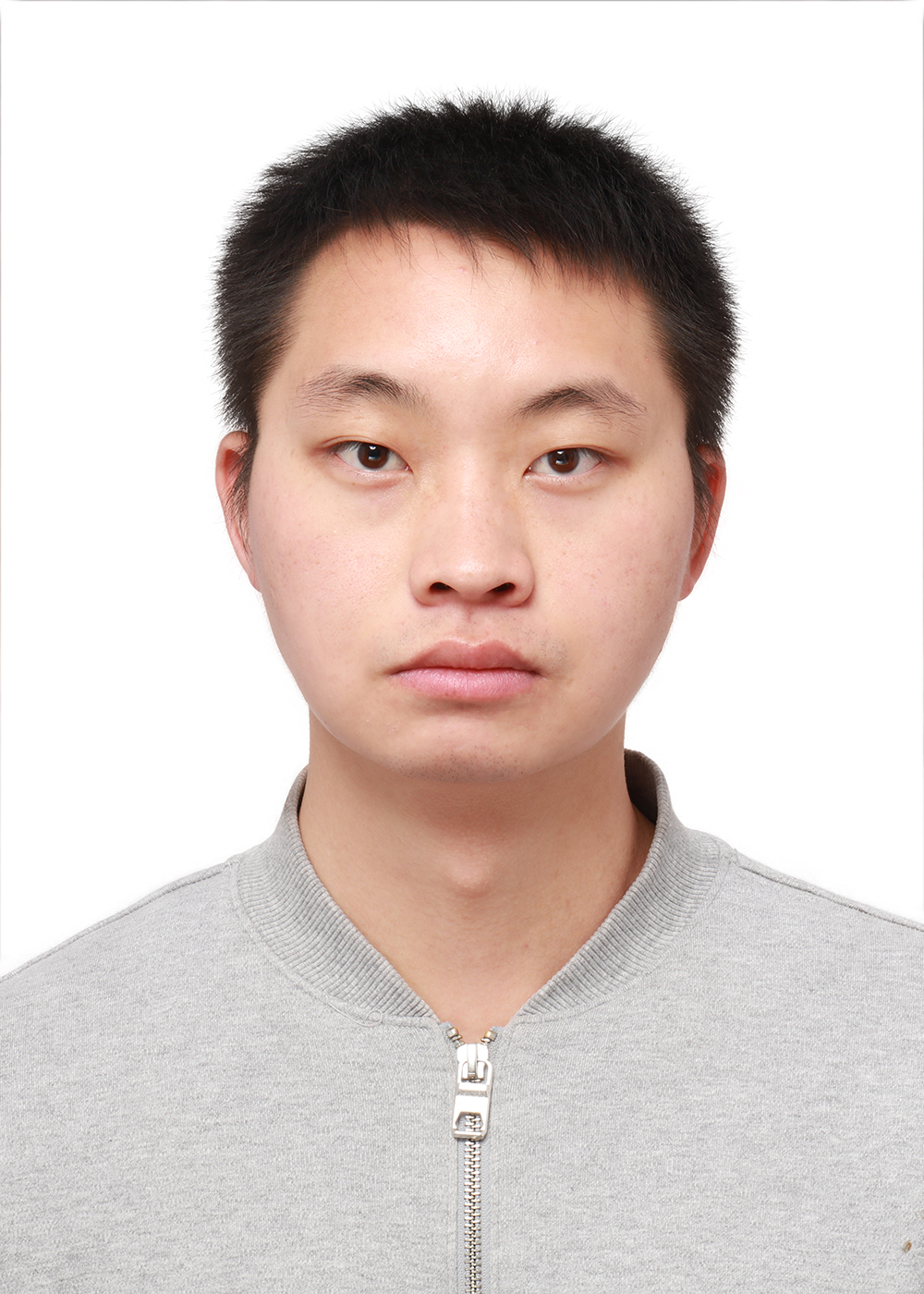}}]{Te Ma}
received the B.E. degree in electronic Information engineering from Chongqin University, Chongqin, China, in 2022. Since 2022, he has been working toward the master's degree with the School of Computer Science and Technology, Xinjiang University, under the supervision of Hao Huang  and Zhijian Ou (during internship at THU-SPMI since January 2023). His research interests include multilingual speech recognition and weakly supervised learning theory.
\end{IEEEbiography}

\newpage

\begin{IEEEbiography}[{\includegraphics[width=1in,height=1.25in,clip,keepaspectratio]{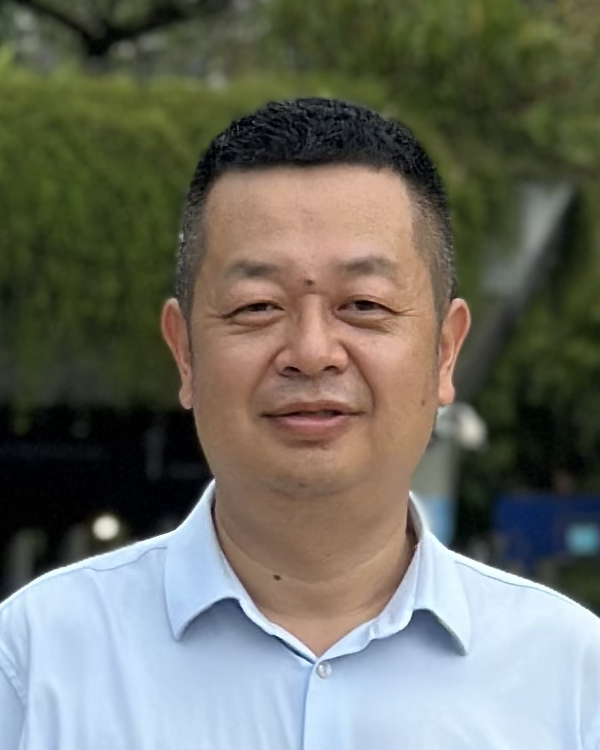}}]{Hao Huang (Member, IEEE)}
received the B.E. degree from Shanghai Jiao Tong University, Shanghai, China, in 1999, the M.E. degree from Xinjiang University, Urumqi, China, 2004, and the Ph.D. degree from Shanghai Jiao Tong University, in 2008, respectively. He is currently a Professor with the School of Computer Science and Technology, Xinjiang University. His current research interests include speech and language processing, and multi-media human-computer interaction. 
\end{IEEEbiography}

\vspace{-13cm}

\begin{IEEEbiography}[{\includegraphics[width=1in,height=1.25in,clip,keepaspectratio]{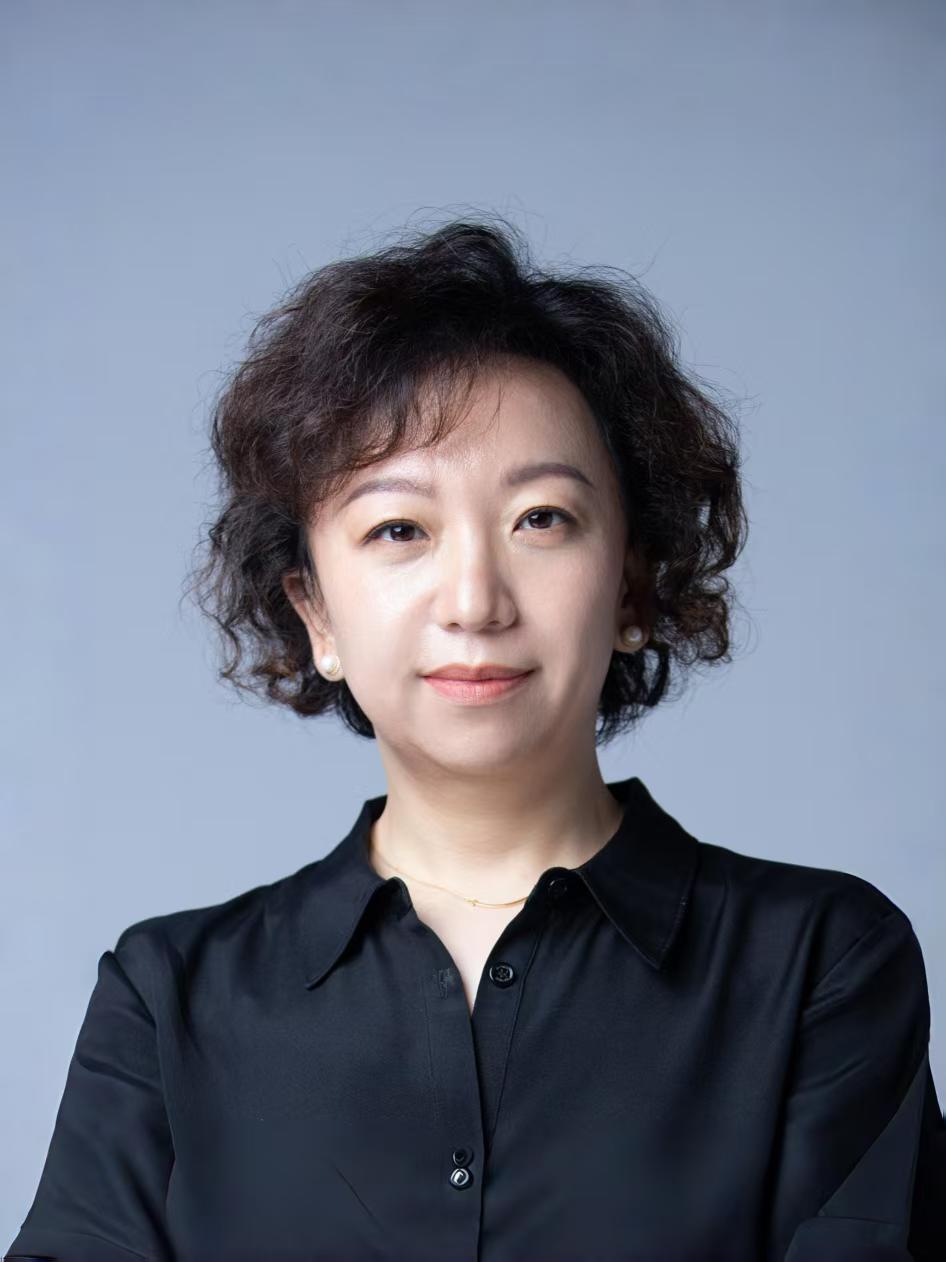}}]{Wenbo Zhao}
received the B.S. degree from Chongqing University of Posts and Telecommunications in, and the M.S. degree from Huazhong University of Science and Technology. She currently serves as the vice president of China Unicom (Guangdong) Industrial Internet Co., Ltd. Her research interests include wireless communications and networking, artificial intelligence, data analytics, and dialogue system, with a particular emphasis on driving the integration of those cutting-edge technologies into next-generation communication systems. She has 3 publications and more than 10 patents.
\end{IEEEbiography}

\vspace{-13cm}

\begin{IEEEbiography}[{\includegraphics[width=1in,height=1.25in,clip,keepaspectratio]{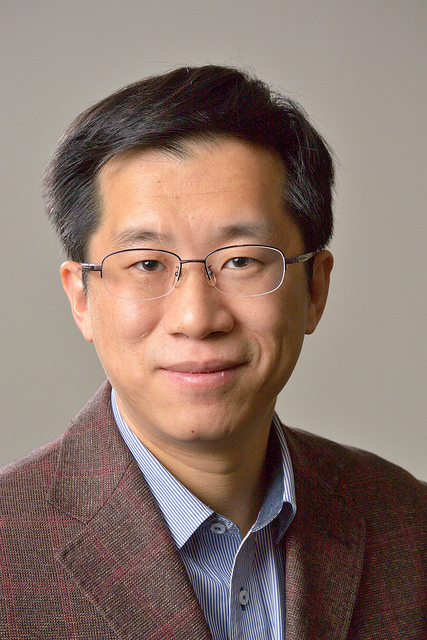}}]{Zhijian Ou (Senior Member, IEEE)}
received the Ph.D. degree from Tsinghua University, Beijing, China, in 2003. He is currently a Professor with the Department of Electronic Engineering, Tsinghua University, and Co-founder of TasiTech. His research interests include speech and language processing (particularly speech recognition and dialogue systems) and machine intelligence (particularly with graphical models and deep learning). He is an Senior Area Editor of IEEE/ACM TRANSACTIONS ON AUDIO, SPEECH AND LANGUAGE PROCESSING and SIGNAL PROCESSING LETTERS, an Editorial Board Member of COMPUTER SPEECH AND LANGUAGE, a Member of IEEE Speech and Language Processing Technical Committee, and was the General Chair of SLT 2021, EMNLP 2022 SereTOD Workshop, SLT 2024 FutureDial-RAG Challenge, and the Tutorial Chair of INTERSPEECH 2020.
\end{IEEEbiography}


\end{document}